\documentclass[prb,showpacs,twocolumn,superscriptaddress,aps,a4paper]{revtex4}
\usepackage{dcolumn}
\usepackage{amsmath}
\usepackage{graphicx}
\usepackage{latexsym}
\usepackage{amsfonts}
\usepackage{amssymb}
\DeclareGraphicsExtensions{.pdf,.gif,.jpg}

\newcommand{\be}{\begin{equation}}
\newcommand{\ee}{\end{equation}}
\newcommand{\beq}{\begin{eqnarray}}
\newcommand{\eeq}{\end{eqnarray}}

\begin{document}

\def\bbe{\mbox{\boldmath $e$}}
\def\bbf{\mbox{\boldmath $f$}}
\def\bg{\mbox{\boldmath $g$}}
\def\bh{\mbox{\boldmath $h$}}
\def\bj{\mbox{\boldmath $j$}}
\def\bq{\mbox{\boldmath $q$}}
\def\bp{\mbox{\boldmath $p$}}
\def\br{\mbox{\boldmath $r$}}
\def\bz{\mbox{\boldmath $z$}}

\def\bfzero{\mbox{\boldmath $0$}}
\def\bfone{\mbox{\boldmath $1$}}

\def\dr{{\rm d}}

\def\tb{\bar{t}}
\def\zb{\bar{z}}

\def\tgb{\bar{\tau}}

\def\bC{\mbox{\boldmath $C$}}
\def\bG{\mbox{\boldmath $G$}}
\def\bH{\mbox{\boldmath $H$}}
\def\bK{\mbox{\boldmath $K$}}
\def\bM{\mbox{\boldmath $M$}}
\def\bN{\mbox{\boldmath $N$}}
\def\bO{\mbox{\boldmath $O$}}
\def\bQ{\mbox{\boldmath $Q$}}
\def\bR{\mbox{\boldmath $R$}}
\def\bS{\mbox{\boldmath $S$}}
\def\bT{\mbox{\boldmath $T$}}
\def\bU{\mbox{\boldmath $U$}}
\def\bV{\mbox{\boldmath $V$}}
\def\bZ{\mbox{\boldmath $Z$}}

\def\bcalS{\mbox{\boldmath $\mathcal{S}$}}
\def\bcalG{\mbox{\boldmath $\mathcal{G}$}}
\def\bcalE{\mbox{\boldmath $\mathcal{E}$}}

\def\bgG{\mbox{\boldmath $\Gamma$}}
\def\bgL{\mbox{\boldmath $\Lambda$}}
\def\bgS{\mbox{\boldmath $\Sigma$}}

\def\bgr{\mbox{\boldmath $\rho$}}
\def\bgs{\mbox{\boldmath $\sigma$}}

\def\a{\alpha}
\def\b{\beta}
\def\g{\gamma}
\def\G{\Gamma}
\def\d{\delta}
\def\D{\Delta}
\def\e{\epsilon}
\def\ve{\varepsilon}
\def\z{\zeta}
\def\h{\eta}
\def\th{\theta}
\def\k{\kappa}
\def\l{\lambda}
\def\L{\Lambda}
\def\m{\mu}
\def\n{\nu}
\def\x{\xi}
\def\X{\Xi}
\def\p{\pi}
\def\P{\Pi}
\def\r{\rho}
\def\s{\sigma}
\def\S{\Sigma}
\def\t{\tau}
\def\f{\phi}
\def\vf{\varphi}
\def\F{\Phi}
\def\c{\chi}
\def\w{\omega}
\def\W{\Omega}
\def\Q{\Psi}
\def\q{\psi}

\def\ua{\uparrow}
\def\da{\downarrow}
\def\de{\partial}
\def\inf{\infty}
\def\ra{\rightarrow}
\def\bra{\langle}
\def\ket{\rangle}
\def\grad{\mbox{\boldmath $\nabla$}}
\def\Tr{{\rm Tr}}
\def\Re{{\rm Re}}
\def\Im{{\rm Im}}
\def\hc{{\rm h.c.}}

\title{Ultrafast manipulation of electron spins in a double quantum
dot device: A real-time view}

\author{Gianluca Stefanucci}
\email{gianluca.stefanucci@roma2.infn.it}
\affiliation{Dipartimento di Fisica, Universit\'a di Roma Tor Vergata,
Via della Ricerca Scientifica 1, I-00133 Rome, Italy}
\affiliation{European Theoretical Spectroscopy Facility (ETSF)}
\affiliation{Laboratori Nazionali di Frascati, Istituto Nazionale di Fisica
Nucleare, Via E. Fermi 40, 00044 Frascati, Italy}

\author{Enrico Perfetto}
\affiliation{Laboratori Nazionali di Frascati, Istituto Nazionale di Fisica
Nucleare, Via E. Fermi 40, 00044 Frascati, Italy}
\affiliation{Dipartimento di Scienza dei Materiali, Universit\'a di
Milano-Bicocca, Via Cozzi 53, 20125 Milano, Italy}

\author{Michele Cini}
\affiliation{Dipartimento di Fisica, Universit\'a di Roma Tor Vergata,
Via della Ricerca Scientifica 1, I-00133 Rome, Italy}
\affiliation{Laboratori Nazionali di Frascati, Istituto Nazionale di Fisica
Nucleare, Via E. Fermi 40, 00044 Frascati, Italy}

\date{\today}

\begin{abstract}

We consider a double quantum dot system with two embedded and
non-aligned spin impurities to manipulate the magnitude and
polarization of the electron spin density. The device is attached
to semi-infinite one-dimensional leads which are treated exactly.
We provide a real-time description of the electron spin dynamics
when a sequence of ultrafast voltage pulses acts on the device.
The numerical simulations are carried out using a spin generalized
and modified version of a recently proposed algorithm for the time
propagation of open systems [Phys. Rev. B {\bf 72}, 035308
(2005)]. Time-dependent spin accumulations and spin currents are
calculated during the entire operating regime which includes spin
injection and read-out processes. The full knowledge of the
electron dynamics allows us to engineer the transient responses
and improve the device performance. An approximate rate equation
for the electron spin is also derived and used to discuss the
numerical results.

\end{abstract}


\maketitle

\section{Introduction}

The ability of controlling magnitude and orientation
of electron spin densities in integrated
molecules and quantum dots
is of utmost importance to bring quantum computation closer to
real life.\cite{nc.2000,asl.2002} The microscopic description of
nanoscale spin devices like, e.g., the two-quantum-bit gate
envisaged by Loss and Di Vincenzo,\cite{ldv.1998} constitutes a
challenging problem in the theory of open systems far from a
steady state. Research activities in the emerging field of
spin-dependent transport\cite{ms.2002} have mainly focussed on
steady-state properties. Only very recently the transient dynamics
of spin polarized currents through quantum dots has attracted some
attention,\cite{mba.2005,fhs.2006,slgj.2007,s.2007} partly due to
experimental advances in manipulating electronic densities with
ultrafast voltage
pulses.\cite{hfcjh.2003,hwvwbek.2003,ehvbwvk.2004,hvbvenkkv.2005,pjtlylmhg.2005,kbtvnmkv.2006,kstbfajfrp.2007}
This paper goes in the same direction and wants to be a further
step toward the bridging of spin dependent transport and
fundamental quantum computation. We perform time-dependent
simulations of the charge and spin dynamics of a nanoscale device
in contact with one-dimensional leads. The semi-infinite leads are
treated {\em exactly}. The results are analyzed within the
framework of non-equilibrium Green's functions.

\begin{figure}[htbp]
\includegraphics[width=.47\textwidth]{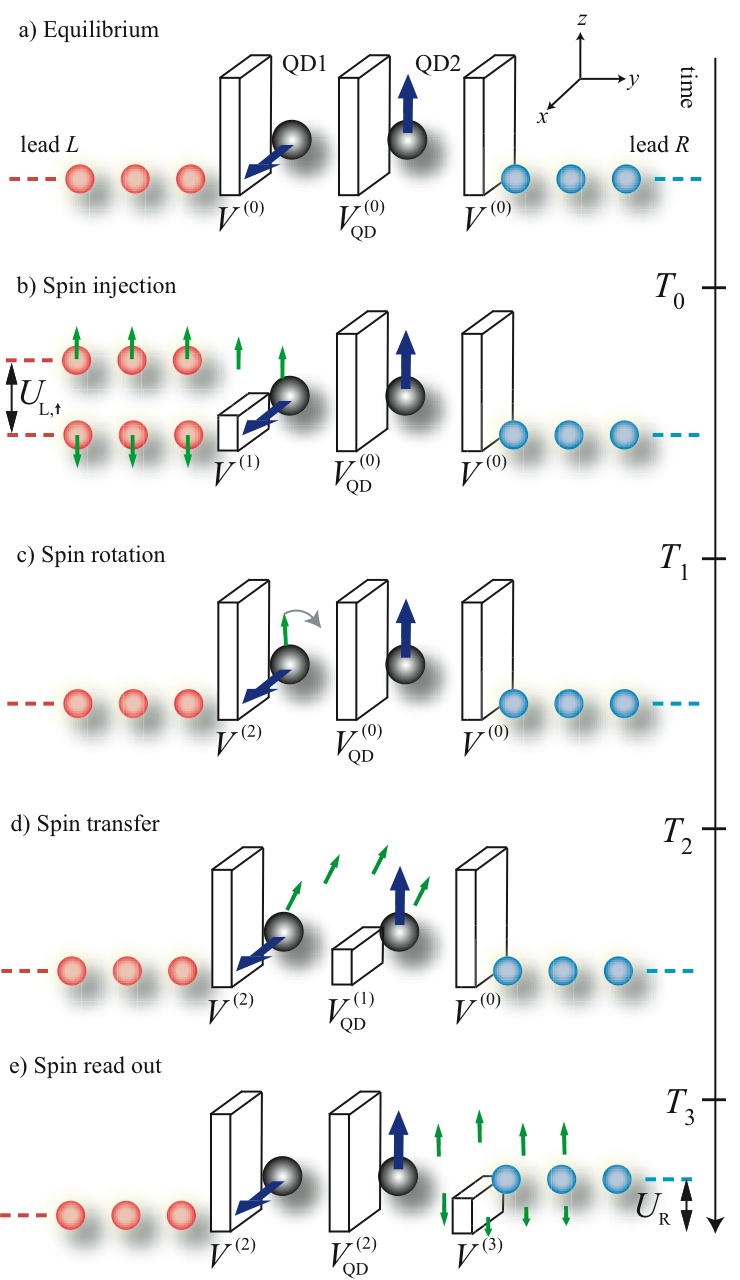}
\caption{Schematic of the double quantum dot device coupled to leads.
At $t<T_{0}=0$ the system is in equilibrium. At $T_{0}$ a spin bias
$U_{L,\ua}$ is switched on and simultaneously the barrier between QD1 and
the left lead is lowered. The injection of spin up electrons ends at
$T_{1}$ when the bias is turned off and the barrier is raised up. Now
the spin in QD1 rotates till $T_{2}$ when the barrier between the dots is
lowered and the electron spin is transfer from QD1 to QD2. At $T_{3}$
the interdot barrier is raised up again while the barrier between QD2 and
the right lead is lowered. Tuning the electrochemical potential in the right
lead $U_{R}$ to be in between the two levels of QD2, we measure a
large (small) spin current if the electron spin is parallel
(antiparallel) to the spin impurity.}
\label{schematic}
\end{figure}

We consider a double quantum dot device to manipulate the spin
orientation of spin-polarized electrons. Both quantum dots contain a
static spin impurity with which the electron spin is coupled.
The exchange coupling constant is much larger than experimentally
accessible Larmor frequencies, a feature that renders the spin
impurity a potentially ultrafast mean to rotate the electron
spin.\cite{cl.2007,kkr.2007}
Model systems of quantum transport through magnetic quantum dots have
been previously used to study the conductance oscillations
of a local nuclear spin in a magnetic field,\cite{zb.2002}
the gauge-invariant nature of the charge and spin conductances,\cite{zc.2005}
the spin-interference and Aharonov-Bohm oscillations in
a quantum ring with embedded magnetic
impurities,\cite{jsj.2001,atz.2005,cpz.2007} and the effects
of the entanglement of two spin impurities
on the conductance.\cite{cpzov.2006,cpzov.2007}

In this paper we focus on the short time response of the system
when subject to a sequence of voltage pulses, as illustrated in
Fig. \ref{schematic}. The injection of spin polarized electrons
from the left lead to the first quantum dot (QD1) is followed by a
rotation of the electron spin in QD1. Afterwards the electron spin
is transferred from QD1 to the second quantum dot (QD2) and its
polarization is maintained parallel/antiparallel to the spin
impurity of QD2. Eventually, the electron spin in QD2 is read out
by calculating the spin current at the interface with the right
lead. We provide a time-dependent description of some crucial
processes in the theory of spin transport, namely the injection of
spins from a lead to a quantum dot and the spin dynamics of a
double quantum dot system weakly coupled to leads. The results of
our analysis include 1) an overshooting of the spin accumulation
during the spin injection phase, 2) a considerable delay in the
spin relaxation for different exchange couplings in QD1 and QD2,
and 3) oscillations in the transient regime whose frequency
depends on the bandwidth of the leads and, therefore, are absent
in the commonly used wide band limit approximation.

The paper is organized as follows. In Section \ref{model} we present
our model system and introduce the basic notation.
A set of approximate equations to describe
different operating processes are derived in
Section \ref{theory}. We obtain a rate equation for the electron
spin of a quantum dot in contact with an electron reservoir and
identify the mechanisms leading to a deterioration of the spin
polarization and to a damping of the spin magnitude.
This analysis will then be used to optimize
the spin injection from one of the leads to one of the quantum dots.
We also investigate the spin transfer between the two quantum
dots for different initial orientations of the electron spin.
In Section \ref{numres} we use a spin-generalized and modified
version of the algorithm of Ref. \onlinecite{ksarg.2005}, see Appendix \ref{alprop},
to perform numerical simulations of the microscopic electron dynamics
of the double quantum dot system. The results are then interpreted
and discussed using the framework developed in Section \ref{theory}.
In Section \ref{conc} we summarize the main findings and discuss
future directions.

\section{Model system}
\label{model}

We consider a basic two spin-impurities model consisting of two
one-dimensional leads coupled to two single-level (per spin)
quantum dots. The first quantum dot is connected to the left lead
($L$), the second quantum dot is connected to the right lead ($R$)
and a hopping term accounts for tunneling of electrons between QD1
and QD2, see Fig. \ref{schematic}. The Hamiltonian describing the
left ($L$) and right ($R$) leads is \beq
H_{\a}&=&\sum_{\s=\ua,\da}\sum_{j=0}^{\inf}\left[
Vc^{\dag}_{j\a,\s}c_{j+1\a,\s}+
V^{\ast}c^{\dag}_{j+1\a,\s}c_{j\a,\s}\right],
\nonumber \\
&+&\sum_{\s=\ua,\da}\ve_{\a,\s}(t)\sum_{j=0}^{\inf}
c^{\dag}_{j\a,\s}c_{j\a,\s} \label{leadham} \eeq with $\a=L,R$. In
Eq. (\ref{leadham}) the quantity $V$ is the hopping integral
between nearest neighbors orbitals and $\ve_{\a,\s}(t)$ is the
time-dependent on-site energy of lead $\a$ which, in general, can
depend on spin. For $\ve_{\a,\s}=0$ the energy window of both $L$
and $R$ continua is $(-2|V|,2|V|)$ and the half-filled system
correspond to a chemical potential $\m_{\rm hf}=0$. The
Hamiltonian of the double quantum dot system reads \beq H_{\rm
QD}&=&\sum_{i=1}^{2}\left[J_{i}\vec{S}_{i}\cdot\sum_{\s\s'}d^{\dag}_{i,\s}\vec{\s}_{\s\s'}d_{i,\s'}
+v_{i}(t)\sum_{\s}d^{\dag}_{i,\s}d_{i,\s}\right]
\nonumber \\
&+&\sum_{\s}\left[V_{\rm QD}(t)d^{\dag}_{1,\s}d_{2,\s}+ V_{\rm
QD}^{\ast}(t)d^{\dag}_{2,\s}d_{1,\s}\right] \label{dqdham} \eeq
with
$\vec{S}_{i}=(\sin\th_{i}\cos\f_{i},\sin\th_{i}\sin\f_{i},\cos\th_{i})$
the spin of the impurity $i=1,2$, $J_{i}>0$ the corresponding
antiferromagnetic exchange coupling, and
$\vec{\s}=(\s_{x},\s_{y},\s_{z})$ the Pauli matrices. The tunnel
barrier between the dots can be tuned by varying an external gate
voltage.\cite{ldv.1998} This is modelled by using time-dependent
gate voltages  $v_{1}(t)$, $v_{2}(t)$ and interdot hopping
integral $V_{\rm QD}(t)$. The first term in Eq. (\ref{dqdham}) is
the Hamiltonian of the two isolated QDs and can conveniently be
rewritten in matrix form as \be
\sum_{i=1}^{2}(d^{\dag}_{i,\ua},d^{\dag}_{i,\da}) \left(
\begin{array}{cc}
    v_{i}+J_{i}\cos\th_{i} &  J_{i}e^{i\f_{i}}\sin\th_{i} \\
    J_{i}e^{-i\f_{i}}\sin\th_{i} & v_{i}-J_{i}\cos\th_{i}
\end{array}
\right)
\left(
\begin{array}{c}
    d_{i,\ua} \\
    d_{i,\da}
\end{array}
\right).
\label{qdh}
\ee
From Eq. (\ref{qdh}) we see that the isolated QD has two
levels at energy $\ve_{i,\pm}=v_{i}\pm J_{i}$. If
$J_{i}>|v_{i}|$ and $\ve_{\a,\s}=0$
one level is above $\m_{\rm hf}$ while the other is below.

The double QD system is connected to the left and
right leads via the tunneling Hamiltonian
\beq
H_{T}&=&\sum_{\s}\left[V_{L}(t)d^{\dag}_{1,\s}c_{0L,\s}
+V_{L}^{\ast}(t)c^{\dag}_{0L,\s}d_{1,\s}\right]
\nonumber \\
&+&
\sum_{\s}\left[V_{R}(t)d^{\dag}_{2,\s}c_{0R,\s}
+V_{R}^{\ast}(t)c^{\dag}_{0R,\s}d_{2,\s}\right].
\eeq
As for the interdot coupling $V_{\rm QD}$ we allow the
hopping integrals $V_{L}(t)$ and $V_{R}(t)$ to be time-dependent.

Below we discuss a sequence of operations to manipulate the
orientation of the electron spin in QD2 for a fixed orientation of
the spin of the injected electrons. Without loss of generality we
assume that for negative times, $t<0$, the whole system is in
equilibrium at chemical potential $\m$ and inverse temperature
$\b$ [Fig.\ref{schematic}a)].\cite{c.1980} The two quantum dots
are initially very weakly coupled to the leads, $V_{\a}<< V$, and
between them, $V_{\rm QD}<<V$. Furthermore, the two energy levels
of both QD1 and QD2 are much larger than the chemical potential,
i.e., $\ve_{i,\pm}>>\m$, and the population on the dots is
practically zero. Starting from this configuration we apply a
sequence of four perturbations to 1) inject spin up electrons on
QD1 [Fig.\ref{schematic}b)] 2) rotate the electron spin in QD1
[Fig.\ref{schematic}c)] 3) transfer the electron spin from QD1 to
QD2 [Fig.\ref{schematic}d)] and 4) read out the polarization of
the electron spin in QD2 [Fig.\ref{schematic}e)]. Due to the wide
range of possible time-dependent perturbations we restrict the
analysis to piece-wise constant (in time) parameters and obtain a
set of approximate equations to study the four different
processes. This study will then help us in selecting the
parameters for target-specific numerical calculations. Full
simulations of the entire sequence will be shown in Section
\ref{numres}.

\section{Theoretical Framework}
\label{theory}

\subsection{Spin injection and spin read-out}

At time $t=0$ we inject spin-up electrons into QD1
by suddenly switching on a spin bias,\cite{ls.2008}
$\ve_{L,\s}(t)=\th(t)U_{L,\s}$, and
reducing the height of the barrier between lead $L$ and QD1,
i.e., $V_{L}(t)=\th(-t)V^{(0)}+\th(t)V^{(1)}$.
The injection process terminates by switching off the spin bias
and raising up the barrier to the equilibrium value $V^{(0)}$. There
are two different mechanisms which contaminate the spin-up injection
with $x$ and $y$ components. The first is the spin-precession around
the spin impurity $\vec{S}_{1}$ while the second is the spin-relaxation due
to the increased electron hopping $V_{L}$. To tackle this problem we
take advantage of the fact that QD1 and QD2 are, during this process,
weakly linked and we only consider the electron dynamics on QD1 in contact
with the left reservoir, i.e., we approximate $V_{\rm QD}=0$.
Using the non-equilibrium Green's function formalism one finds the
following equation for the lesser Green function
$G^{<}_{\s\s'}$ on QD1
\beq
i\frac{\dr}{\dr t}\bG^{<}(t;t)=
[\bH_{\rm QD1},\bG^{<}(t;t)]
+\int_{0}^{\inf}\dr \tb
\nonumber \\
\times \left[ \bgS_{L}^{<}(t;\tb)\bG^{\rm A}(\tb;t) +\bgS_{L}^{\rm
R}(t;\tb)\bG^{<}(\tb;t) +\hc \right], \label{g<1} \eeq where we
use boldface to indicate $2\times 2$ matrices in spin space and
the symbol ``$[,]$'' denotes a commutator. In the above equation
$\bH_{\rm QD1}=v_{1}+J_{1}\vec{S}_{1}\cdot\vec{\s}$ is the
one-particle Hamiltonian of the isolated QD1 while $\bgS_{L}$ is
the embedding self-energy of lead $L$. We have discarded the
integral between 0 and $-i\b$ along the imaginary time axis since
$V_{L}(t<0)=V^{(0)}<<V$ and hence $\bgS_{L}\sim 0$ in
equilibrium.\cite{sa.2004} The superscripts R/A in $\bgS_{L}$ and
$\bG$ denote retarded/advanced components. The self-energy is
diagonal in spin space since there is no spin-flip hopping between
lead $L$ and QD1. In terms of one-particle eigenstates $\q_{k}(j)$
and eigenenergies $\ve_{k}$ of lead $L$ one finds for $t,t'>0$
\beq \S_{L,\s\s'}^{\rm R}(t;t')&=&\d_{\s\s'}
|V^{(1)}|^{2}\int\frac{\dr \w}{2\p} e^{-i(\w+U_{L,\s})(t-t')}
\nonumber \\
&\times& \sum_{k}
\frac{|\q_{k}(0)|^{2}}{\w-\ve_{k}+i\eta},
\label{sigr1}
\eeq
\beq
\S_{L,\s\s'}^{<}(t;t')&=&\d_{\s\s'}
|V^{(1)}|^{2}\int\frac{\dr \w}{2\p}
e^{-i(\w+U_{L,\s})(t-t')}if(\w)
\nonumber \\
&\times& 2\p\sum_{k}|\q_{k}(0)|^{2}\d(\w-\ve_{k}).
\label{sig<1}
\eeq
At low temperatures and low biases only frequencies close to the Fermi
energy $\ve_{\rm F}$ are probed. Using the Wide Band Limit (WBL)
approximation, i.e.,
\be
\sum_{k}|\q_{k}(0)|^{2}\d(\w-\ve_{k})\sim \r_{\rm F}
\ee
with $\r_{\rm F}=\r(\ve_{\rm F})$ the local density of states at the
interface, we can approximate Eqs. (\ref{sigr1}-\ref{sig<1}) as
\be
\S_{L,\s\s'}^{\rm R}(t;t')=-\frac{i}{2}\d_{\s\s'}\G\d(t-t'),
\ee
\beq
\S_{L,\s\s'}^{\rm <}(t;t')=i\d_{\s\s'}
\G\int\frac{\dr \w}{2\p}e^{-i(\w+U_{L,\s})(t-t')}f(\w)
\nonumber \\
\simeq
-\d_{\s\s'}\frac{\G}{2\p}e^{-i\ve_{\rm F,\s}(t-t')}
\left[\frac{1}{t-t'}-i\p\d(t-t')\right],
\label{sig<2}
\eeq
with $\G=2\p|V^{(1)}|^{2}\r_{\rm F}$ and
$\ve_{\rm F,\s}=\ve_{\rm F}+U_{L,\s}$. In Eq. (\ref{sig<2}) we have
further approximated $\bgS^{<}_{L}$ with its expression at zero
temperature.

Inserting these results into Eq. (\ref{g<1}) one obtains
\beq
i\frac{\dr}{\dr t}\bG^{<}(t;t)=
[\bH_{\rm QD1},\bG^{<}(t;t)]
-\frac{i}{2}\{\bgG,\bG^{<}(t;t)\}
\nonumber \\ -\bgG
-\left[\frac{\bgG}{2\p}\int_{0}^{t}\dr \tb\,
\frac{\exp[-i\bcalE_{\rm F}(t-\tb)]}{t-\tb}
\bG^{\rm A}(\tb;t)+\hc\right],
\label{g<2}
\eeq
where the symbol ``$\{,\}$'' denotes the anticommutator and
the matrices $[\bgG]_{\s\s'}=\d_{\s\s'}\G$ and
$[\bcalE_{\rm F}]_{\s\s'}=\d_{\s\s'}\ve_{\rm F,\s}$. From Eq.
(\ref{g<2}) we can extract a rate equation for the electron spin
\be
\vec{S}_{1,\rm el}(t)\equiv -\frac{i}{2}\Tr\left[\bG^{<}(t;t)\vec{\s}\right]
\ee
on QD1. In the WBL approximation the advanced Green's function reads
\beq
\bG^{\rm
A}(\tb;t)&=&i\theta(-\D t)\exp[-i(\bH_{\rm QD1}+\frac{i}{2}\bgG)\D t]
\nonumber \\
&=&i\theta(-\D t)e^{-i(v_{1}+\frac{i}{2}\G)\D t}
\nonumber \\
&\times&
\left[
\cos(J_{1}\D t)-i\sin(J_{1}\D t)\vec{S}_{1}\cdot\vec{\s}
\right],
\eeq
with $\D t\equiv \tb-t$.
Substituting this result into Eq. (\ref{g<2}), multiplying with
$\vec{\s}$ and tracing over the spin indices we find
\beq
\frac{\dr}{\dr t}\vec{S}_{1,\rm el}&=&
J_{1}(\vec{S}_{1}\wedge \vec{S}_{1,\rm el})-\frac{\G}{2}\vec{S}_{1,\rm el}
\nonumber \\
&-&
\frac{\G}{\p}\int_{0}^{t}\frac{\dr \tb}{\D t}\,
e^{\frac{\G}{2}\D t}\cos((\ve_{+}-v_{1})\D t)
\nonumber \\
&\times&
[ \cos(\ve_{-}\D t)\sin(J_{1}\D t)\vec{S}_{1}
\nonumber \\
&-&\sin(\ve_{-}\D t)\{\cos(J_{1}\D t)\hat{z}
-\sin(J_{1}\D t)\hat{z}\wedge \vec{S}_{1}\}],
\nonumber \\
\label{sre} \eeq with $\hat{z}$ the unit vector in the $z$
direction, and $\ve_{\pm}=\ve_{\rm F,\ua}\pm\ve_{\rm F,\da}$. It
is instructive to expand the right hand side in powers of $t$. To
first order in $t$ one finds a simplified rate equation for the
electron spin \be \frac{\dr}{\dr t}\vec{S}_{1,\rm el}=
J_{1}(\vec{S}_{1}\wedge \vec{S}_{1,\rm
el})-\frac{\G}{2}\vec{S}_{1,\rm el}
+\frac{t}{\p}\left(\G\ve_{-}\hat{z} -J_{1}\G\vec{S}_{1}\right),
\label{fot} \ee which is reliable for times $t^{-1}\geq
\max[\ve_{-},J_{1}]/2\p$. From the above equation we can identify
four different contributions. The term proportional to $\hat{z}$
is the spin-injection term and is responsible for an increase of
the electron spin along the $z$ direction. Such increase is
quadratic in time and faster the larger the difference
$\ve_{-}=U_{L,\ua}-U_{L,\da}$ is. The first and the last terms are
responsible for a deterioration of the spin direction due to spin
precession (first term) and spin relaxation (last term). The
latter drives the electron spin towards a configuration
antiparallel to the spin impurity $\vec{S}_{1}$. Finally, the
second term is responsible for an overall damping of the spin
magnitude. Going beyond the first order in $t$, see Eq.
(\ref{sre}), one observes the appearance of a new relaxation
direction, that is $\hat{z}\wedge \vec{S}_{1}$. This latter result
is completely general as it is only dictated by the symmetry of
the system.

We wish to emphasize that the rate equation (\ref{sre}) has been
derived under the sole assumption that the quantum dot QD1 is initially
isolated and then contacted with lead $L$. This is the same
situation occurring in the spin read-out phase when the barrier
between the weakly coupled QD2 and lead $R$ is lowered. Thus, the
rate equation for $\vec{S}_{2,\rm el}$ during the read-out phase is
identical to Eq.  (\ref{sre}) for $\vec{S}_{1,\rm el}$ even though the
parameters are different and, more importantly, different initial
conditions must be imposed.

\subsection{Spin rotation and spin transfer}
\label{stottra}

After a time $T_{1}$ the spin bias is switched off and the hopping
$V_{L}$ is again reduced to values much smaller than $V$. In this
phase QD1 is well isolated and the electron spin precesses around
the spin impurity $\vec{S}_{1}$ according to
\be
\frac{\dr}{\dr t}\vec{S}_{1,\rm el}=
J_{1}(\vec{S}_{1}\wedge \vec{S}_{1,\rm el}),\quad
t>T_{1}.
\label{srp}
\ee

Let us now specialize to the situation illustrated in Fig.
\ref{schematic}
with $\vec{S}_{1}$ oriented along the positive $x$ axis and $\vec{S}_{2}$ along the
positive $z$ axis. We recall that the Fermi energy is much smaller than
the energy levels of the two isolated quantum dots and hence
that the equilibrium electron density is
vanishingly small.
For $J_{1}>>\ve_{-}$, see Eq. (\ref{fot}), we expect an efficient
injection of spin up electrons in QD1 and for $J_{1}>>\G$
a major contamination along the $y$ direction. This implies that the electron
spin $\vec{S}_{1,\rm el}(T_{1})$ has a  small $x$ component at the end of
the spin injection process. Since $\vec{S}_{1}$ is parallel to the $x$
axis $\vec{S}_{1,\rm el}(t)$ rotates
in the $yz$ plane for $t>T_{1}$. We let the system evolve till a time
$T_{2}>T_{1}$ and we approximate $\vec{S}_{1,\rm el}(T_{2})=
(0,S_{1,\rm el}^{y}(T_{2}),S_{1,\rm el}^{z}(T_{2}))$ on the $yz$ plane and  $\vec{S}_{2,\rm el}(T_{2})=0$
(this latter approximation comes from the fact that
$V_{\rm QD}$ and $V_{R}$ are both much smaller than $V$ for $t<T_{2}$).

At $t=T_{2}$ we transfer the electron spin by lowering
the barrier between QD1 and QD2. This corresponds to an increase
of the interdot hopping $V_{\rm QD}$.
Letting $|\F(T_{2})\ket$ be the evolved many-particle state of the
entire system at $t=T_{2}$,
the density matrix $\bgr$
of the double quantum dot system has matrix elements
$[\bgr]_{i\s,j\s'}=\bra
\F(T_{2})|d^{\dag}_{j,\s'}d_{i,\s}|\F(T_{2})\ket$.
It is convenient to introduce the notation
$1=(1,\ua)$, $2=(1,\da)$, $3=(2,\ua)$ and
$4=(2,\da)$ for the collective index $(i,\s)$. The density matrix is
then represented by the following $4\times 4$ matrix
\be
\bgr=2S_{1,\rm el}(T_{2})\left(
\begin{array}{cccc}
    \cos^{2}\th & -\frac{i}{2}\sin2\th & 0 & 0 \\
    \frac{i}{2}\sin2\th & \sin^{2}\th & 0 & 0 \\
    0 & 0 & 0 & 0 \\
    0 & 0 & 0 & 0
\end{array}
\right),
\label{rt2}
\ee
with $\sin2\th=S_{1,\rm el}^{y}(T_{2})/S_{1,\rm el}(T_{2})$,
$\cos 2\th=S_{1,\rm el}^{z}(T_{2})/S_{1,\rm el}(T_{2})$, and
$S_{1,\rm el}=\sqrt{\vec{S}_{1,\rm el}\cdot \vec{S}_{1,\rm el}}$ the
spin magnitude. We are
interested in how to choose the
angle $\th$ in order to maximize the electron spin of QD2 along the $z$
direction (parallel/antiparallel to the spin impurity $\vec{S}_{2}$).
For simplicity we take the gate voltages $v_{1}=0$ and $v_{2}=0$. Then
the isolated double quantum dot system is described by the
$4\times 4$ Hamiltonian matrix
\be
\bH_{\rm QD}=
\left(\begin{array}{cc}
J_{1}\s_{x} & V_{\rm QD}\bfone_{2} \\
V_{\rm QD}\bfone_{2} & J_{2}\s_{z}
\end{array}
\right),
\ee
with $\bfone_{2}$ the $2\times 2$ identity matrix. In terms of
$\bgr$ and $\bH_{\rm QD}$ the $z$ component $S^{z}_{2,\rm el}$ of the
electron spin on QD2 is
given by
\be
S^{z}_{2,\rm el}(t+T_{2})=
\frac{1}{2}\Tr\left[
\bgS_{2}^{z}\exp(i\bH_{\rm QD}t)\bgr
\exp(-i\bH_{\rm QD}t)\right],
\ee
with the spin operator of QD2
\be
\bgS_{2}^{z}=\left(
\begin{array}{cc}
    \bfzero_{2} & \bfzero_{2} \\
    \bfzero_{2} & \s_{z}
\end{array}
\right),
\ee
and $\bfzero_{2}$ the $2\times 2$ null matrix. Substituting $\bgr$ from
Eq. (\ref{rt2}) we find
\beq
\frac{S^{z}_{2,\rm el}(t+T_{2})}{S_{1,\rm el}(T_{2})}
&=&\frac{i}{2}\sin2\th
\left\{
[\bgS_{2}^{z}(t)]_{1,2}-[\bgS_{2}^{z}(t)]_{2,1}
\right\}
\nonumber \\
&+&\cos^{2}\th[\bgS_{2}^{z}(t)]_{1,1}
+\sin^{2}\th[\bgS_{2}^{z}(t)]_{2,2},
\nonumber \\
\eeq
where we have defined the spin operator in the Heisenberg
representation
\be
\bgS_{2}^{z}(t)\equiv
\exp(-i\bH_{\rm QD}t)\bgS_{2}^{z}\exp(i\bH_{\rm QD}t).
\ee
It is easy to prove that the function $O(t)\equiv
\frac{i}{2}([\bgS_{2}^{z}(t)]_{1,2}-[\bgS_{2}^{z}(t)]_{2,1})$ is an odd function of
time while $[\bgS_{2}^{z}(t)]_{1,1}$ and $[\bgS_{2}^{z}(t)]_{2,2}$
are even functions of time. In appendix \ref{proof}
we further prove that
\be
E(t)\equiv[\bgS_{2}^{z}(t)]_{1,1}=-[\bgS_{2}^{z}(t)]_{2,2},
\label{ir}
\ee
which leads to the simple formula
\beq
S^{z}_{2,\rm el}(t+T_{2})&=&S_{1,\rm el}(T_{2})
[O(t)\sin2\th +
E(t)\cos 2\th]
\nonumber \\
&=&O(t)S^{y}_{1,\rm el} +
E(t)S^{z}_{1,\rm el}.
\eeq

The function $E(t)$ can be written as a linear combination of the
cosine functions $\cos(\w_{\m\n}t)$ while $O(t)$ as a linear
combination of the sine functions $\sin(\w_{\m\n}t)$, where
$\w_{\m\n}=\ve_{\m}-\ve_{\n}$ is the difference between two
eigenvalues of $\bH_{\rm QD}$. The eigenvalues $\ve_{\m}$,
$\m=1,\ldots,4$ can be calculated analytically and read \be
\ve_{\m}=\pm \sqrt{\frac{J_{+}^{2}+2V_{\rm
QD}^{2}\pm\sqrt{J_{-}^{4}+4J_{+}^{2}V_{\rm QD}^{2}}}{2}},
\label{2dqd} \ee with $J^{2}_{\pm}=J_{1}^{2}\pm J_{2}^{2}$.

\begin{figure}[htbp]
\includegraphics*[width=.47\textwidth,height=5.5cm]{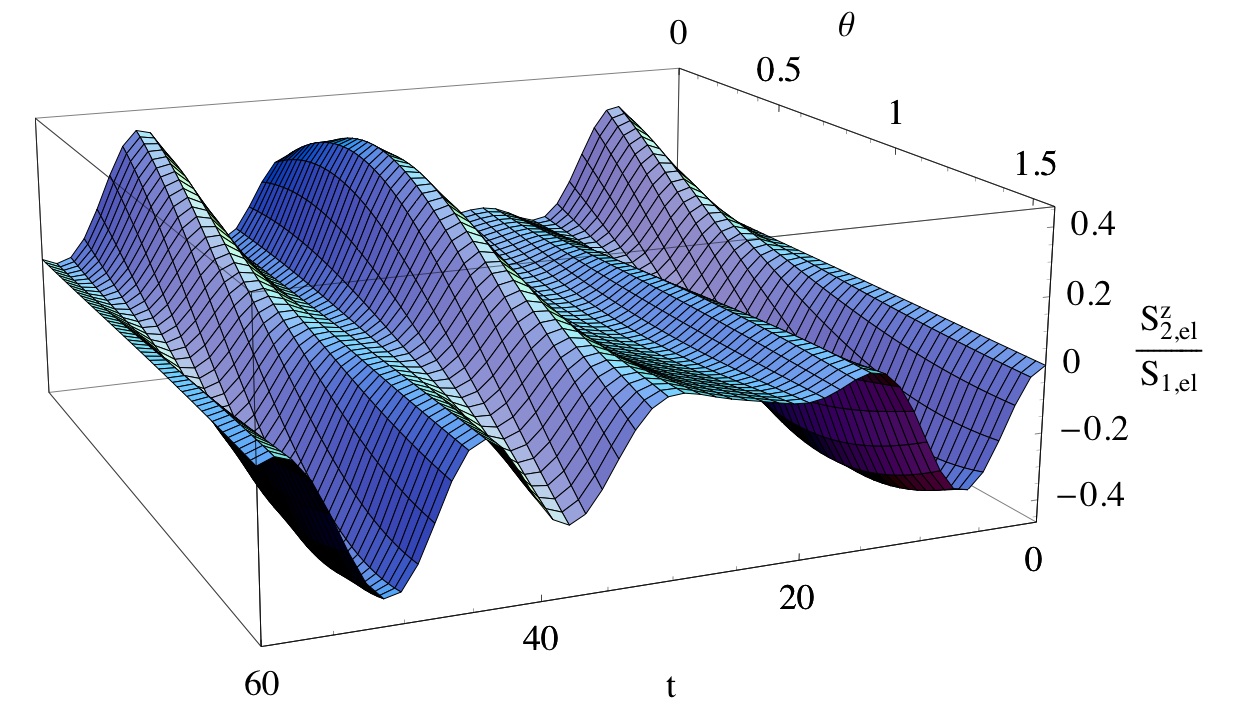}
\caption{Ratio $S^{z}_{2,\rm
el}(t+T_{2})/S_{1,\rm el}(T_{2})$ versus time $t$ and
initial polarization $\th$ for $J_{1}=J_{2}=0.1$ and
$V_{\rm QD}=0.2$.}
\label{figana1}
\end{figure}

As an example, in Fig. \ref{figana1} we plot the ratio $S^{z}_{2,\rm
el}(t+T_{2})/S_{1,\rm el}(T_{2})$ as a function of time $t$ and
initial polarization $\th$ for $J_{1}=J_{2}=0.1$ and
$V_{\rm QD}=0.2$. We notice that for most polarizations
$S^{z}_{2,\rm el}(t+T_{2})$ remains smaller than $0.2$. Only for some
special value of $\th$ the $z$ component of the spin in QD2 reaches a value
larger than 0.4. This means that the maximum efficiency
in transferring an electron spin polarized in the $yz$ plane from QD1 to QD2
with final polarization along the positive $z$ axis is about
$80-90\%$.

All above processes can be numerically simulated
without resorting to any of the approximations employed in
this Section. This allows for a more quantitative investigation of the
device performance and will be the topic of the next Section.

\section{Numerical simulations and discussion}
\label{numres}

In this Section we use a modified version of the
algorithm proposed in Ref. \onlinecite{ksarg.2005}
to propagate in time finite systems in contact with infinitely long
leads, see Appendix \ref{alprop}, and
investigate the microscopic dynamics of the spin-injection, the
spin-accumulation as well as the spin rotation of conducting
electrons scattering against the double QD device of Eq. (\ref{dqdham}).
In the following analysis energies are measured in units of $V$,
times in units of $\hbar/V$, spins in units of $\hbar$ and currents
in units of $eV/\hbar$, with $e$ the electron charge.
The full Hamiltonian is time
independent for negative times and the system is in equilibrium
at zero temperature and Fermi energy $\ve_{\rm F}$.\cite{c.1980} We
start by considering two identical QDs with exchange coupling
$J_{1}=J_{2}=0.1$ and gate potential $v_{1}=v_{2}=0$ weakly coupled
to the left and right leads, $V_{L}=V_{R}=V^{(0)}=0.01$, and with interdot
hopping $V_{\rm QD}=0.01$.
Choosing $V\sim 100$ meV the exchange couplings $J_{1},J_{2}\sim 10$ meV lie in
the physical parameter range\cite{kkr.2007} and the corresponding time unit is
$\hbar/V \sim 100$ fs which is appropriate to study ultrafast
dynamics.\cite{gksa.2001,mklsa.2007}
The impurity spin $\vec{S}_{1}$ of QD1 is oriented
along the positive $x$ axis while $\vec{S}_{2}$ is oriented along the
positive $z$ axis. The on-site energies of the leads $\ve_{\a,\s}$
are initially all zero.

\subsection{Spin injection}

At time $t=0$ we switch on a spin bias
$\ve_{L,\ua}(t)=\th(t)U_{L,\ua}$ in lead $L$ for spin up electrons
 ($U_{L,\da}=0$) and increase the hopping $V_{L}$ from $V^{(0)}$ to $V^{(1)}$.

\begin{figure}[htbp]
\includegraphics*[width=.47\textwidth,height=5.5cm]{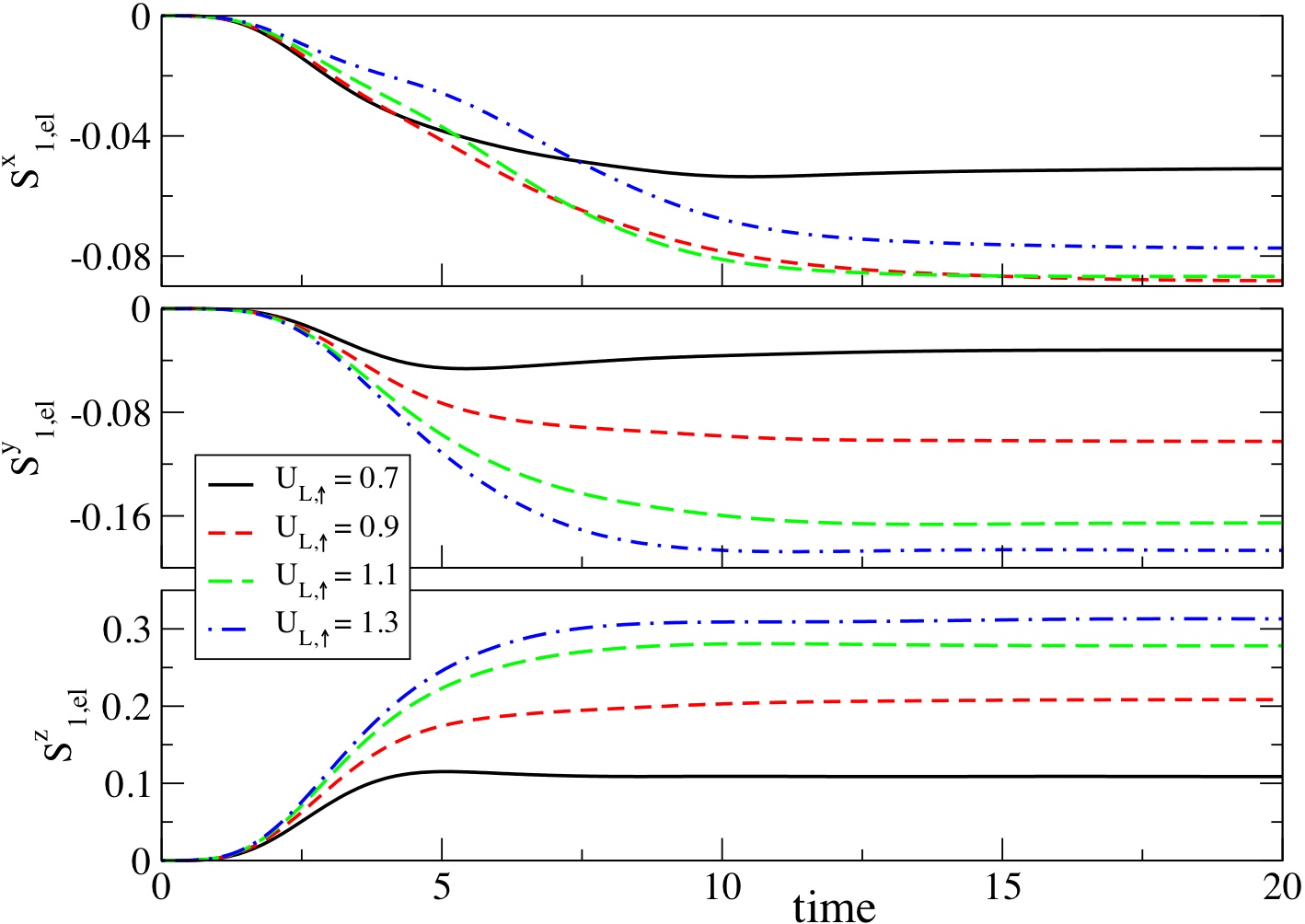}
\caption{The three components of the electron spin in QD1 versus time
for a sudden change of the hopping $V_{L}$ from $V^{(0)}=0.01$ to
$V^{(1)}=0.5$ and a simultaneous sudden switching of the spin-bias
$U_{L,\ua}=0.7,\;0.9,\;1.1,\;1.3$. The equilibrium parameters
are $J_{1}=J_{2}=0.1$, $V_{L}=V_{R}=V_{\rm QD}=V^{(0)}$ and the Fermi
energy is $\ve_{\rm F}=0.96$.}
\label{spinbias}
\end{figure}

In Fig. \ref{spinbias} we study the spin-injection process when the
Fermi energy is $\ve_{\rm F}=-0.96$ (which correspond to an initial
electron occupation on QD1 of the order of $10^{-5}$)
and the hopping between $L$ and QD1 at
positive times is $V^{(1)}=0.5$.
We calculate the time-dependent
expectation value of the spin of the conducting electrons
$\vec{S}_{1,{\rm el}}$
on QD1 for different biases $U_{L,\ua}$. Since
$U_{L,\ua}\sim 1 > J_{1}$ the rate equation (\ref{fot}) is reliable
for times $t\leq 2\p/U_{L,\ua}\sim 2\p$. In this time window we
observe that the $z$ component $S_{1,{\rm el}}^{z}$ increases
quadratically in time and that the rate is larger the larger
is the spin bias $U_{L,\ua}$, in agreement with Eq. (\ref{fot})
(we recall that in this case $\ve_{-}=U_{L,\ua}$).
The $y$ component $S_{1,{\rm el}}^{y}$ has a trend similar to
$S_{1,{\rm el}}^{z}$
but the transient is even smoother. This can be
explained by observing that as the spin up electrons enter QD1 they undergo
a spin rotation due to the spin impurity oriented
along the positive $x$ axis. Taking into account that for small $t$
we have $S_{1,{\rm el}}^{z}\sim t^{2}$, from Eq. (\ref{fot}) we see that
$S_{1,{\rm el}}^{y}\sim J_{1}t^{3}$. As the $z$ component also the
$x$ component $S_{1,{\rm el}}^{x}$ grows quadratically in time.
From Eq. (\ref{fot}) one finds $S_{1,{\rm el}}^{z}(t)/S_{1,{\rm
el}}^{x}(t)\sim \ve_{-}/J_{1}$ meaning that to minimize the
contamination of spin up electrons with an $x$ component
it must be $\ve_{-}/J_{1}>1$.

\begin{figure}[htbp]
\includegraphics*[width=.47\textwidth,height=5.5cm]{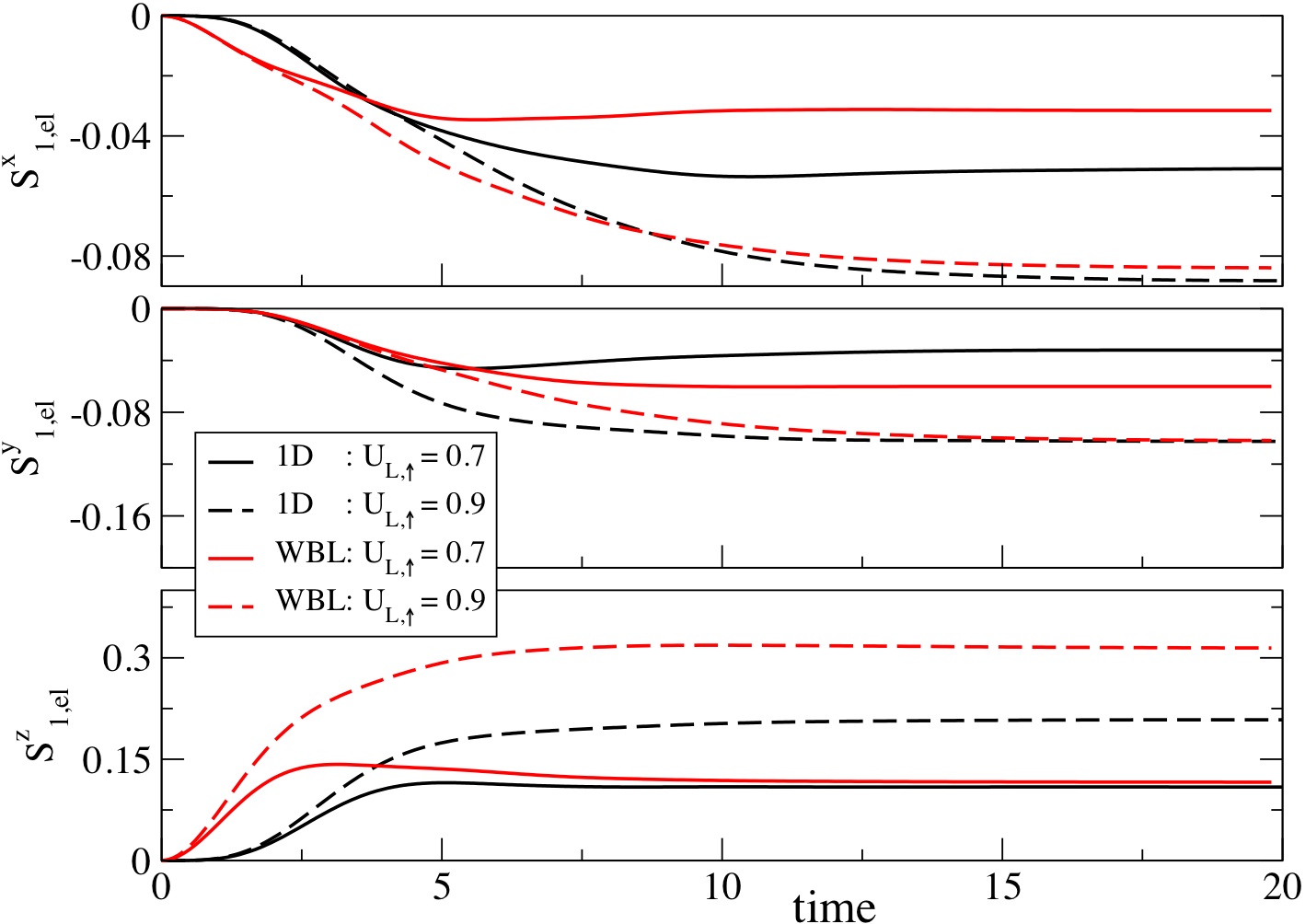}
\caption{Comparison between the results obtained with one-dimensional
(1D)
leads and WBL leads for the electron spin in QD1 and
$U_{L,\ua}=0.7,\;0.9$. The other parameters are the same as in Fig. \ref{spinbias}.}
\label{wideband}
\end{figure}

We wish to observe that at intermediate biases $U_{L,\ua}$ the
numerical results agree with the rate equation (\ref{sre}) only qualitatively.
The comparison between the time evolution of the electron spin
in QD1 for $U_{L,\ua}=0.7,\;0.9$ as obtained with one-dimensional leads
and with leads treated  in the WBL approximation is shown in Fig. \ref{wideband}.

\begin{figure}[htbp]
\includegraphics*[width=.47\textwidth,height=5.5cm]{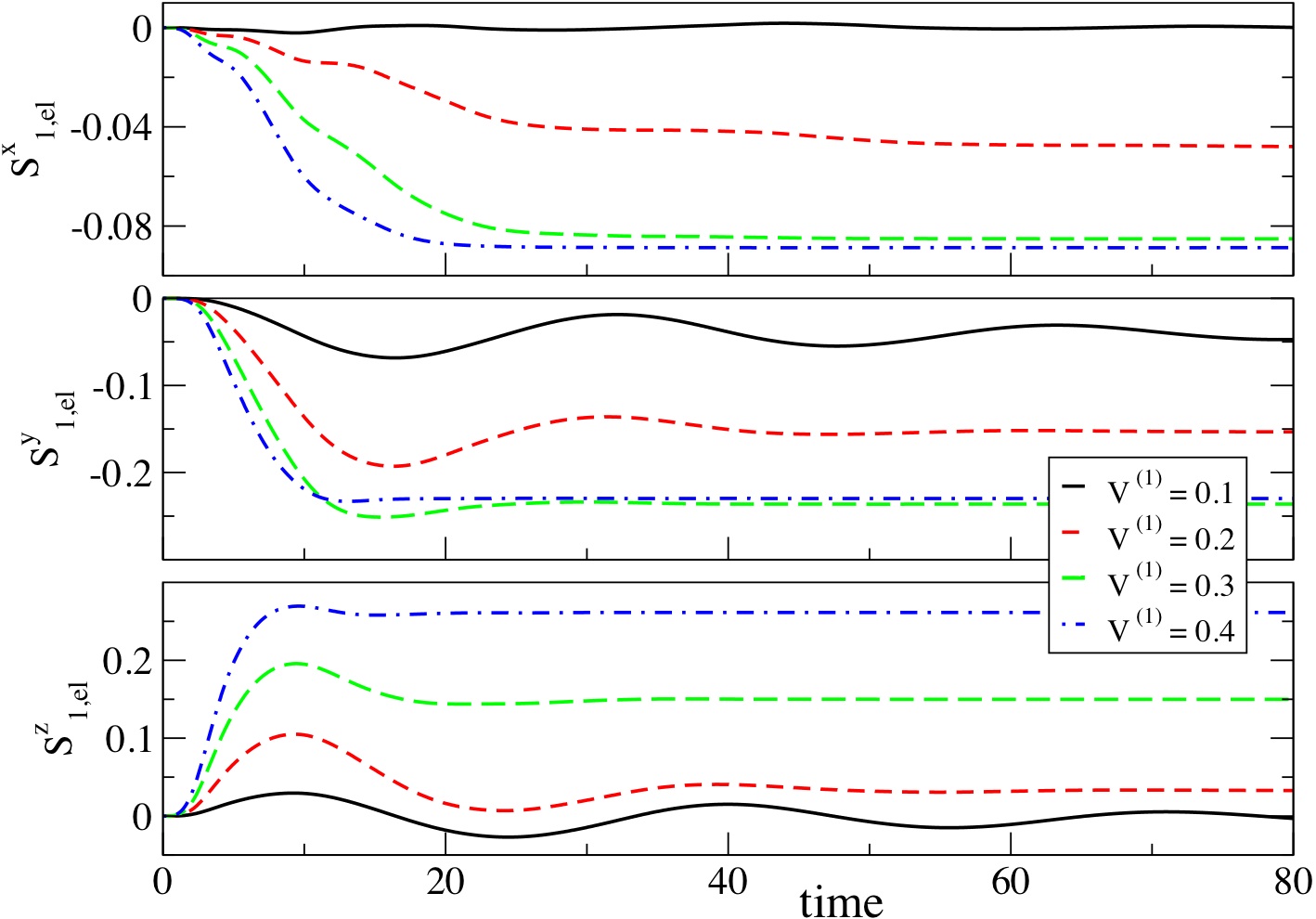}
\caption{The three components of the electron spin in QD1 versus time
for a sudden switching of the spin-bias $U_{L,\ua}=1.3$ and a
simultaneous sudden change of the hopping $V_{L}$ from $V^{(0)}=0.01$ to
$V^{(1)}=0.1,\;0.2,\;0.3,\;0.4$. The other parameters are the same as in
Fig. \ref{spinbias}.}
\label{spinhop}
\end{figure}

In Fig. \ref{spinhop} we fix the bias for spin-up electrons to be
$U_{L,\ua}=1.3$ and analyze the spin-dynamics on QD1 for different values
of $V^{(1)}$. We first observe that
the transient time decreases by increasing $V^{(1)}$ and hence $\G$.
This is easily understood by noticing
that the second term in
Eq. (\ref{sre}) yields an exponential damping of the spin oscillations.
The spin oscillations can be
observed in the $y$ and $z$ components and are due to the spin
precession around the spin impurity $\vec{S}_{1}$.
The period of the oscillation is $T=\p/J_{1}$ and is independent of $V^{(1)}$,
as it should. It is also interesting to
observe that for small times $S_{1,{\rm el}}^{z}$ {\em overshoots its
steady value}
and hence more efficient spin injections may be achieved by
properly engineering the transient response. In our case,
for an efficient spin up injection only the ratios $r_{xy}=|S_{1,{\rm
el}}^{x}/S_{1,{\rm el}}^{y}|$ and $r_{xz}=|S_{1,{\rm
el}}^{x}/S_{1,{\rm el}}^{z}|$ must be small at the end of the process
since the $y$ component can be reduced to zero in the second phase
when $V_{L}<<V$  and $\vec{S}_{1,\rm el}$ can precess around the spin impurity.
From Fig. \ref{spinhop} we find that for $V_{\rm par}=0.2$ and at $t\sim 10$ the ratios
$r_{xy}\sim 0.1$ and $r_{xz}\sim 0.12$ while at the steady state
$r_{xy}\sim 0.31$ and $r_{xz}\sim 1.43$.

\subsection{Spin rotation}

\begin{figure}[htbp]
\includegraphics*[width=.47\textwidth,height=5.5cm]{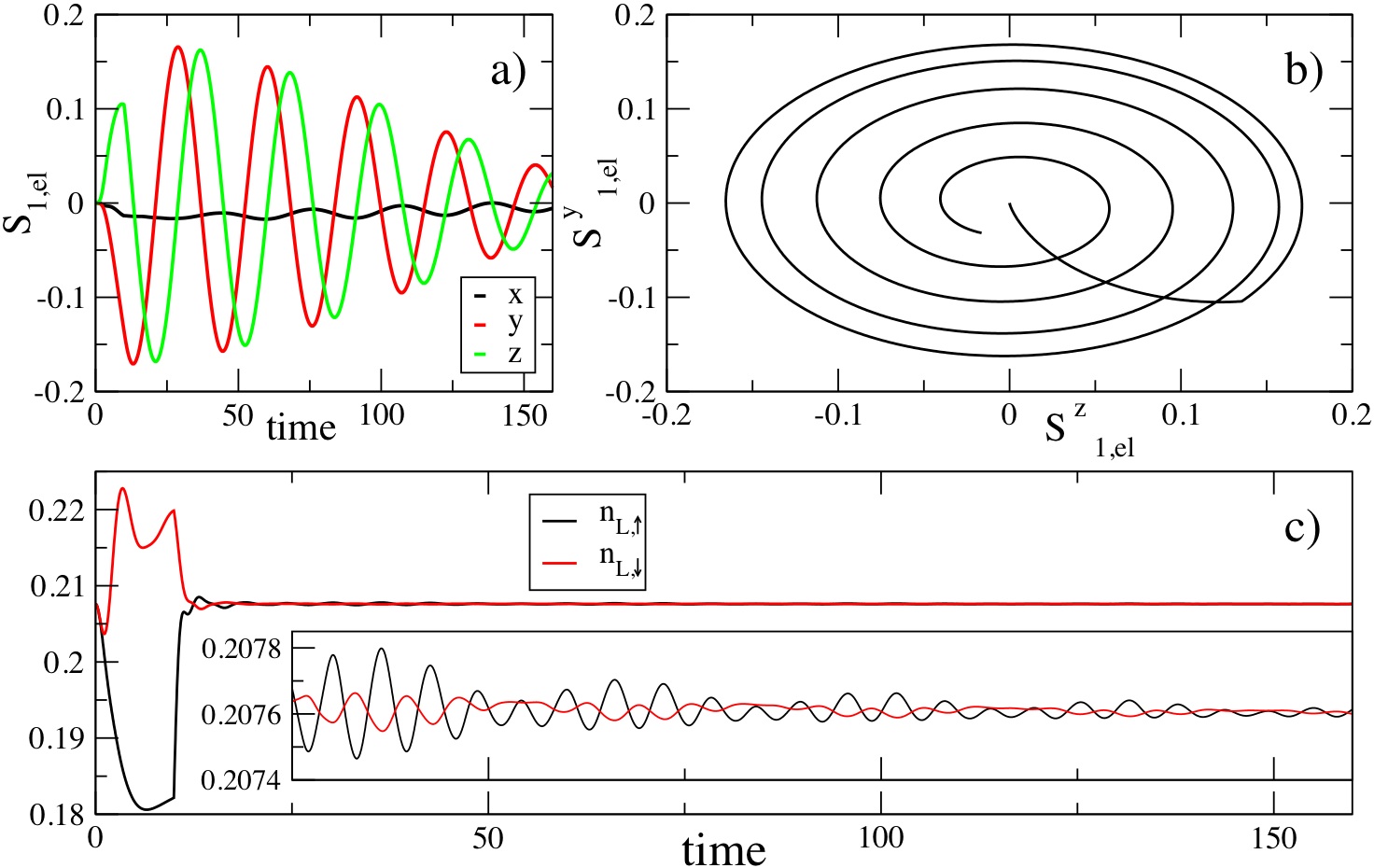}
\caption{Electron spin in QD1 and electron density at the left
interface during the injection phase
($t<T_{1}$) and the rotation phase ($t>T_{1}$) when $T_{1}=10$. The
equilibrium parameters are the same as in Fig. \ref{spinhop}, i.e., 
$J_{1}=J_{2}=0.1$, $V_{L}=V_{R}=V_{\rm QD}=V^{(0)}$ and the Fermi
energy is $\ve_{\rm F}=0.96$. For
$0<t<T_{1}$ the system is perturbed by a spin bias $U_{L,\ua}=1.3$
and a larger hopping $V_{L}=V^{(1)}=0.2$. At $t=T_{1}$ the spin bias
is switched off and the hopping $V_{L}$ is suddenly changed to
$V^{(2)}=0.01$. Panel a) The three components of $\vec{S}_{1,\rm
el}$. Panel b) The trajectory of the projection of $\vec{S}_{1,\rm
el}$ onto the $yz$ plane. Panel c) Spin up and down density on the
first site of the left lead. The inset is a magnification of both
densities $n_{L,\ua}$ and $n_{L\da}$ for times $t>25$.}
\label{rotV001bis}
\end{figure}

The injection process terminates after some time $T_{1}$ by
switching the spin bias off and raising back the barrier between
QD1 and the left lead, i.e., $V_{L}(t)=\th(-t)V^{(0)}+
\th(t)\th(T_{1}-t)V^{(1)}+\th(t-T_{1})V^{(2)}$. During the second
phase QD1 is weakly coupled to the environment and the electron
spin precesses around $\vec{S}_{1}$. Let us focus on the situation
discussed above with $V^{(1)}=0.2$ and let $T_{1}=10$ be the
duration of the first phase. In Fig. \ref{rotV001bis} we study the
electron spin on QD1 [panels a) and b)] and the densities
$n_{L,\s}\equiv\bra c^{\dag}_{0L,\s}c_{0L,\s}\ket$ on the first
site of the left electrode [panel c)] for $V^{(2)}=0.01$. The
contaminating component  $S_{1,\rm el}^{x}$ ceases to decrease at
$t=T_{1}$ while the $y$ and $z$ components are well described by
damped cosine functions with a phase lag of $\p/(4J_{1})$ [panel
a)]. Due to the weak contact $V^{(2)}$ the magnitude $S_{1,\rm
el}=\sqrt{\vec{S}_{1,\rm el}\cdot \vec{S}_{1,\rm el}}$ of the
electron spin changes on a time scale much longer than the
spin-exchange time-scale $\sim 1/J_{1}$. This is shown in panel b)
where the trajectory of $\vec{S}_{1,\rm el}$ is projected onto the
$yz$ plane. For times $t<T_{1}$ the trajectory has a large radial
component while for $t>T_{1}$ the spin moves along a spiral
trajectory. It is also interesting to look at the densities on the
nearest neighbor site of QD1 [panel c)]. During the first phase
($t<T_{1}$) a majority of spin up electrons are transferred from
lead $L$ to QD1 and, as a consequence, $n_{L,\ua}$ decreases. On
the contrary the density $n_{L,\da}$ increases due to the
following two-step mechanism. As the spin up electrons hop from
lead $L$ to QD1 they undergo a spin rotation and acquire a down
component. These electrons have about zero energy and can easily
hop to the left lead where the spin-down band is filled up to
$\ve_{\rm F}+U_{L,\da}=\ve_{\rm F}=-0.96$. At the end of the
injection process the densities change abruptly and approach their
initial value since $V^{(2)}=V^{(0)}$. The inset of panel c) is a
magnification of the curves $n_{L,\ua}(t)$ and $n_{L,\da}(t)$ for
$25<t<160$. It is clearly visible a quantum beating in both
densities due to the alignment of the spin impurity along the $x$
axis. In both cases two oscillations with frequency $|\ve_{\rm
F}\pm J_{1}|= 0.96\pm 0.1$ are superimposed to an envelope
oscillation of frequency $2J_{1}=0.2$.

\begin{figure}[htbp]
    \vspace{0.2cm}
\includegraphics*[width=.47\textwidth,height=5.5cm]{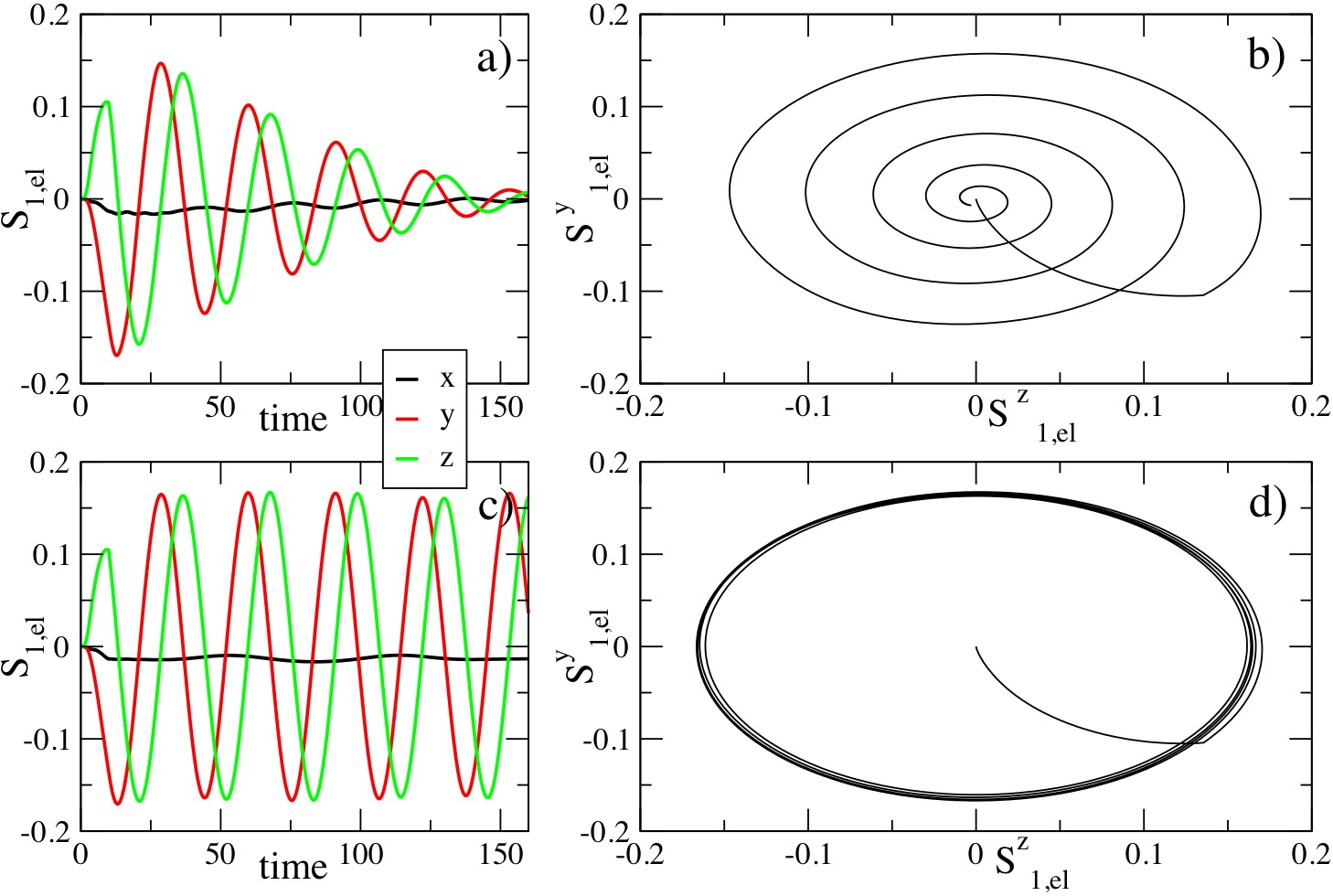}
\caption{The three components of $\vec{S}_{1,\rm el}$ [panel a)] and
the trajectory of the projection of $\vec{S}_{1,\rm el}$ onto the $yz$ plane
[panel b)] for the same system as in Fig. \ref{rotV001bis} except that
the hopping parameter $V^{(2)}=0.06$ is six times larger.
Panels c)-d) like panels a)-b) but the equilibrium parameter
$J_{2}=0.02$ while the hopping parameter $V^{(2)}=0.01$.}
\label{vj}
\end{figure}

The spin rotation phase is further investigated in Fig. \ref{vj}
where we consider the same system as in Fig. \ref{rotV001bis} except for
the value of the hopping parameter $V^{(2)}=0.06$ which is six times larger
[panels a) and b)] or the exchange coupling $J_{2}=0.02$ which is
five times smaller [panels c) and d)].
In the first case the $x$ component remains an order of magnitude smaller than
$S_{1,\rm el}$ [see panel a)] and eventually approaches a steady
value slightly larger than the initial one [not shown].
As in Fig. \ref{rotV001bis} the electron spin is damped in all three
directions but it decays faster. The projection of $\vec{S}_{1,\rm el}$
onto the $yz$ plane [panel b)] yields a spiral trajectory which finishes
very close to the origin after a time $t\sim 160$.
On the other hand, {\em for a smaller coupling $J_{2}=J_{1}/5$ we do not
appreciate any damping} within the time propagation window $t<160$.
The $y$ and $z$ components of $\vec{S}_{1,\rm el}$ are well described by two
undamped cosine functions with a phase lag $\p/(4J_{1})$ and an
amplitude which is about ten times larger than $|S^{x}_{1,\rm el}|$,
see panel c). In panel d) we show the projection of $\vec{S}_{1,\rm el}$ onto
the $yz$ plane. The reduced damping is a desirable feature and
has to be attributed to the
mismatch of the energy levels in the two quantum dots: $\pm J_{1}$ in
QD1 and $\pm J_{2}$ in QD2.

\begin{figure}[htbp]
    \vspace{0.2cm}
\includegraphics*[width=.235\textwidth,height=4.5cm]{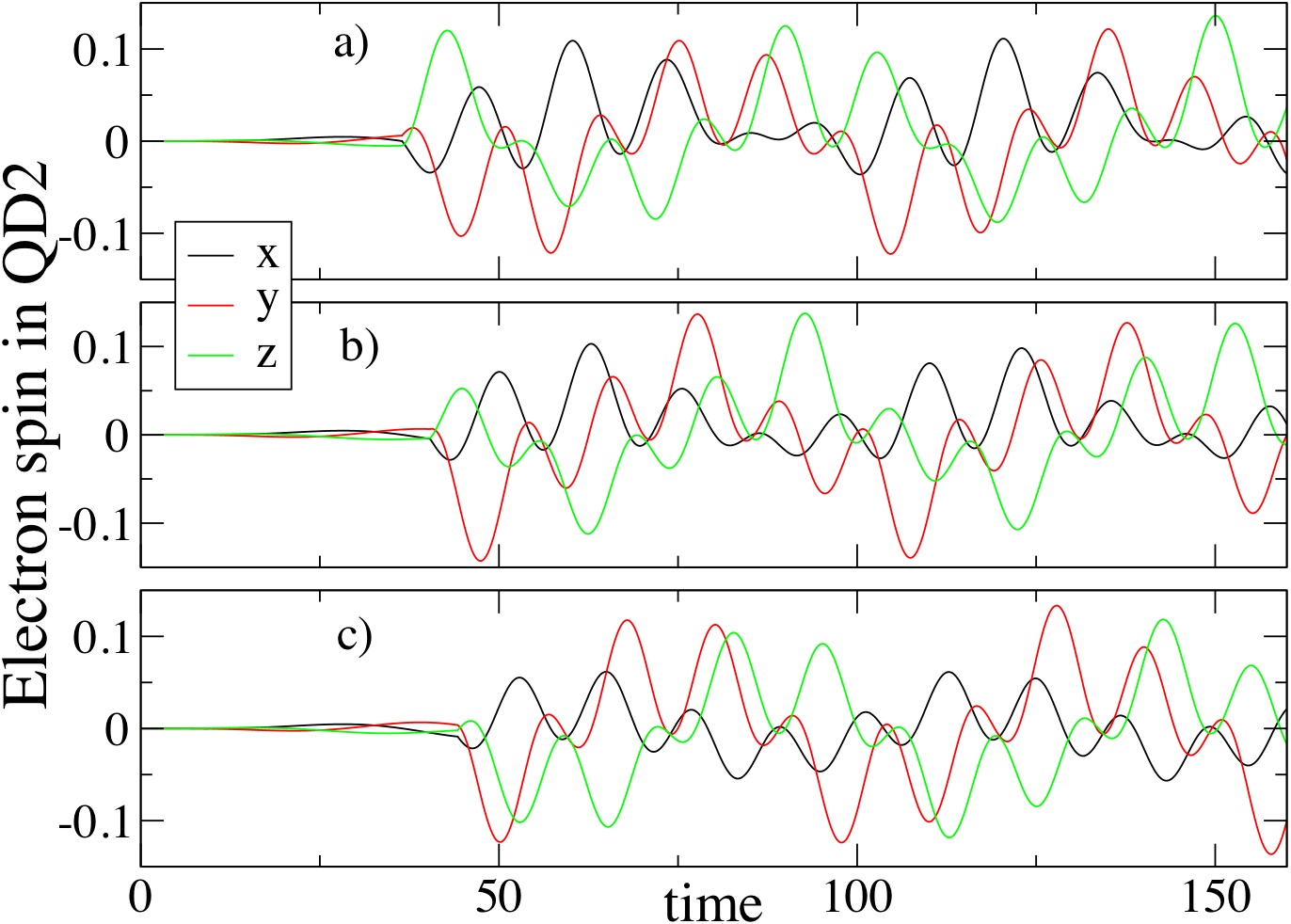}
\includegraphics*[width=.235\textwidth,height=4.5cm]{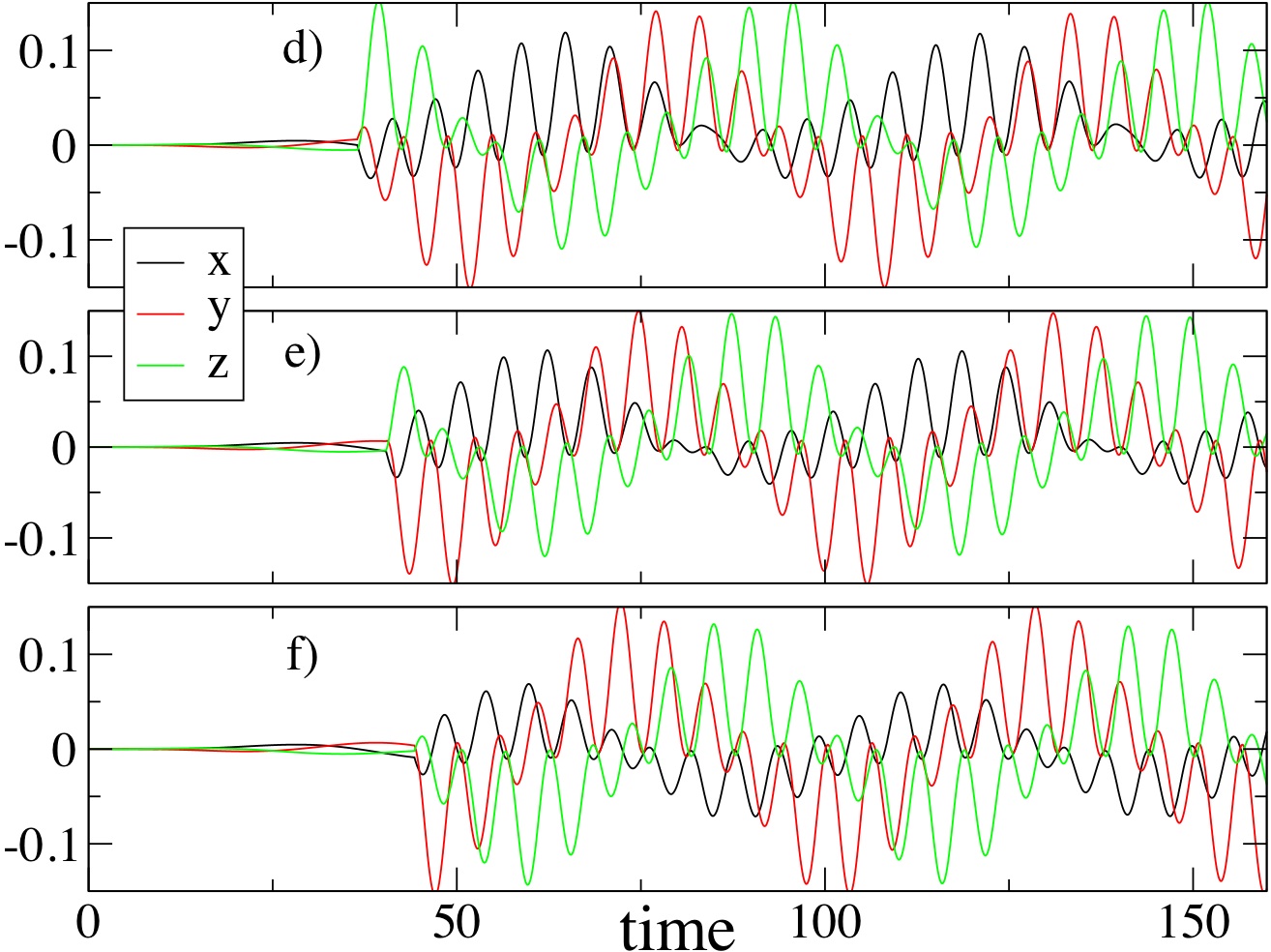}
\caption{The electron spin $\vec{S}_{2,\rm el}$ in QD2 before
($t<T_{2}$) and after ($t>T_{2}$) the spin transfer phase when
the interdot hopping is set to $V_{\rm QD}^{(1)}=0.2$ [panels a) to c)]
and $V_{\rm QD}^{(1)}=0.5$ [panels d) to f)]. The transfer phase
starts at  $T_{2}\sim 36.5$ [panels a) and d)], $T_{2}\sim 40.4$ [panels b)
and e)], and $T_{2}\sim 44.3$ [panels c) and f)]. All parameters
before time $T_{2}$ are the same as in Fig. \ref{rotV001bis}
except for the exchange coupling $J_{2}=0.05$.}
\label{vqd}
\end{figure}

\subsection{Spin transfer}

The rotation of the electron spin in QD1 ($t>T_{1}$) terminates
at $t=T_{2}>T_{1}$ when the barrier between
QD1 and QD2 is lowered and, as a consequence, the interdot hopping
increases, i.e., $V_{\rm QD}=V_{\rm QD}^{(0)}\th(T_{2}-t)+V_{\rm
QD}^{(1)}\th(t-T_{2})$, where $V_{\rm QD}^{(0)}=0.01$.
This is the spin transfer phase. In Fig. \ref{vqd} we plot the three
components of the electron spin in QD2 versus time for
$V_{\rm QD}^{(1)}=0.2$ [panels a) to c)] and $V_{\rm QD}^{(1)}=0.5$
[panels d) to f)]. For times $t<T_{2}$ the system undergoes the same
perturbations as in Figs. \ref{rotV001bis}-\ref{vj}.
Here we have considered an exchange
coupling in QD2 of $J_{2}=0.05$ and a hopping $V^{(2)}=0.01$.
The frequency of the oscillations is larger the
larger the interdot coupling is, in agreement with Eq. (\ref{2dqd}).
The efficiency of the transfer has been investigated for different
times $T_{2}$ at which the electron spin in QD1 is polarized
along $\hat{z}$ [$T_{2}\sim 36.5$, panels a) and d)],
$\frac{1}{\sqrt{2}}(\hat{z}-\hat{y})$ [$T_{2}\sim 40.4$, panels b)
and e)], and $-\hat{y}$ [$T_{2}\sim 44.3$, panels c) and f)]. For our
choice of parameters the efficiency is higher if the spin in QD1 is
polarized along $\hat{z}$.

\begin{figure}[htbp]
    \vspace{0.2cm}
\includegraphics*[width=.235\textwidth,height=4.5cm]{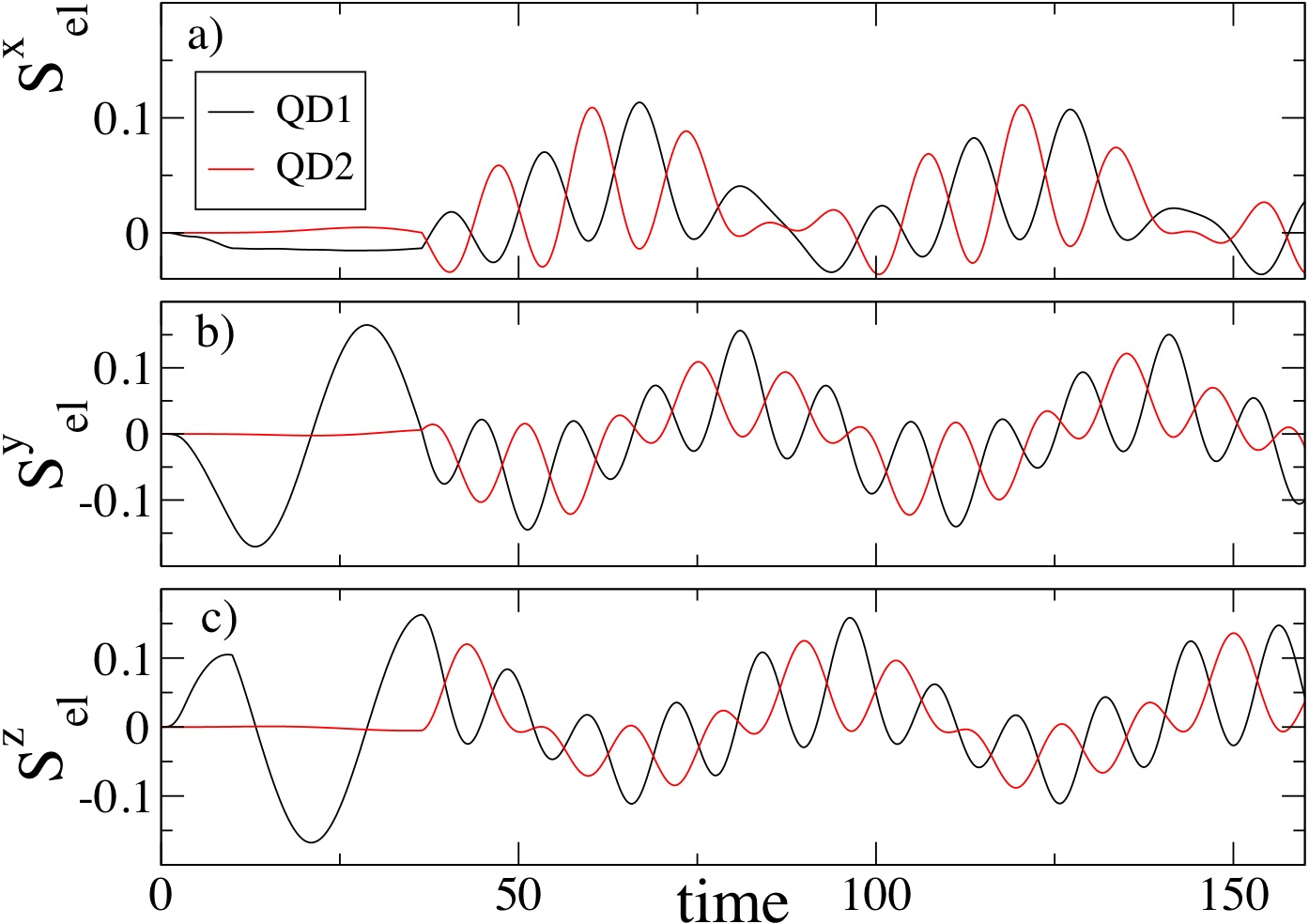}
\includegraphics*[width=.235\textwidth,height=4.5cm]{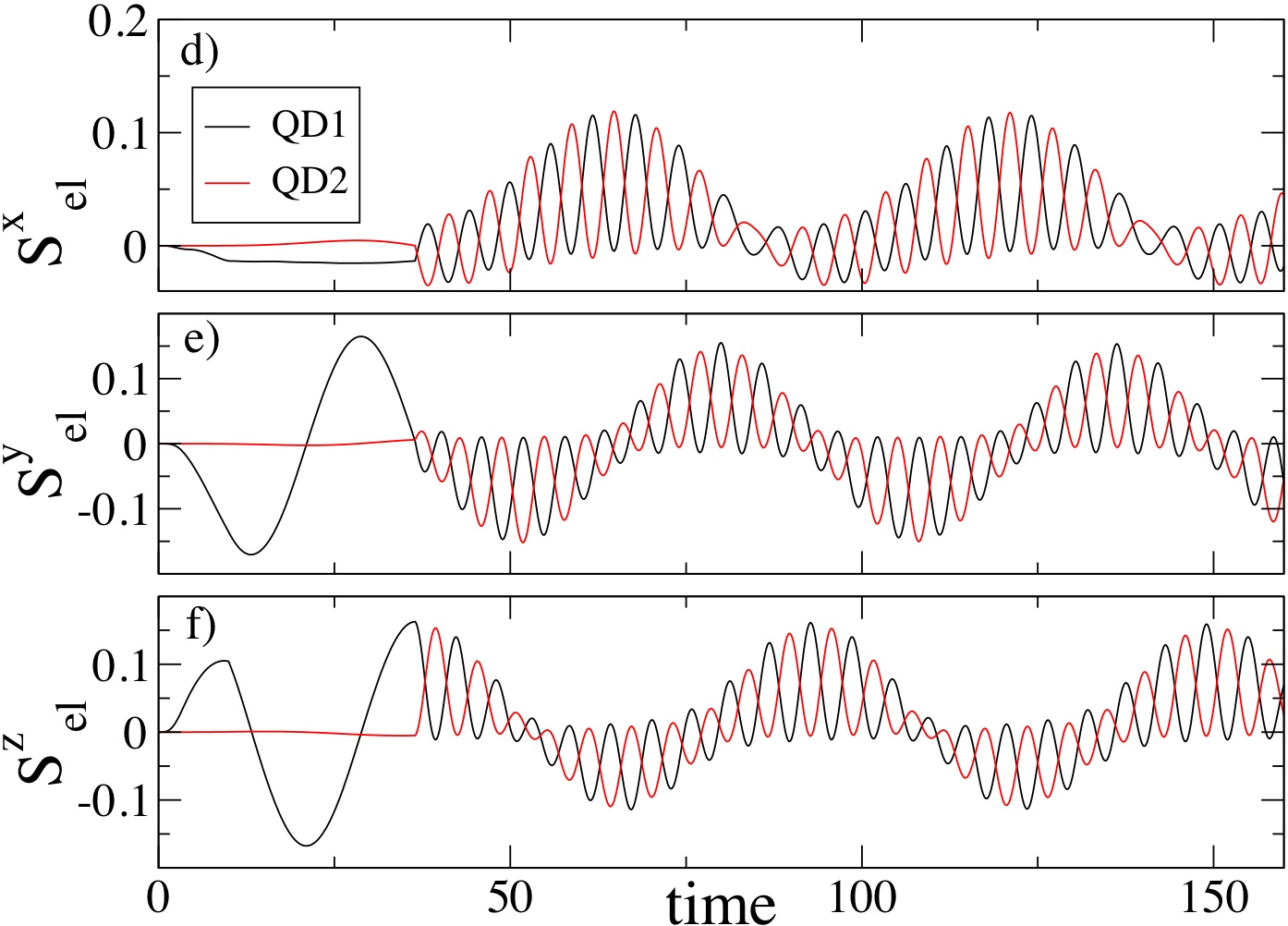}
\caption{Electron spin in QD1 and QD2 for $T_{2}=36.5$ and $V_{\rm
QD}^{(1)}=0.2$ [panels a) to c)] and $V_{\rm QD}^{(1)}=0.5$
[panels d) to f)]. The other parameters are the same as in Fig.
\ref{vqd}.} \label{vqdsz}
\end{figure}

We also observe that for all three
components the maxima of the electron spin in QD1
correspond to the minima of the electron spin in QD2,
see Fig. \ref{vqdsz} where we plot $\vec{S}_{1,\rm el}$ and
$\vec{S}_{2,\rm el}$ for $T_{2}=36.5$ and $V_{\rm QD}^{(1)}=0.2$ [panels a) to c)]
and $V_{\rm QD}^{(1)}=0.5$ [panels d) to f)].
From Fig. \ref{vqd} panel d) and Fig. \ref{vqdsz} panels d) to f)
one observes that when $T_{2}$ corresponds to the time at which
$\vec{S}_{1,\rm el}(T_{2})$ is polarized along $\hat{z}$, the
maxima of $S^{z}_{2,\rm el}$ are close to the zeros of
$S^{x}_{2,\rm el}$ and $S^{y}_{2,\rm el}$, in agreement
with the analysis of Section \ref{stottra}.
We define the ratio $r_{\perp}\equiv
S^{z}_{2,\rm el}/\sqrt{(S^{x}_{2,\rm el})^{2}+(S^{y}_{2,\rm el})^{2}}$.
In the propagation window the maxima of $S^{z}_{2,\rm el}$ occur at
$t=39.32$ when $r_{\perp}\sim 0.28$ and $S^{z}_{2,\rm el}=0.153$,
$t=95.68$ when $r_{\perp}\sim 0.26$ and $S^{z}_{2,\rm el}=0.152$,
and $t=151.96$ when $r_{\perp}\sim 0.24$ and $S^{z}_{2,\rm el}=0.151$.
Taking into account that $S^{z}_{1,\rm el}(T_{2})=0.163$ the
efficiency of the spin transfer can be up to $90\%$.

\subsection{Spin read out}

At a time $t=T_{3}$ when $S^{z}_{2,\rm el}$ has a maximum or a minimum,
the interdot hopping is lowered, i.e., $V_{\rm QD}(t)=
V_{\rm QD}^{(0)}\theta(T_{2}-t)+V_{\rm
QD}^{(1)}\theta(t-T_{2})\th(T_{3}-t)+V_{\rm QD}^{(2)}\theta(t-T_{3})$
with $V_{\rm QD}^{(2)}<< V$, and the spin transfer
phase ends.

\begin{figure}[htbp]
    \vspace{0.2cm}
\includegraphics*[width=.47\textwidth,height=5.5cm]{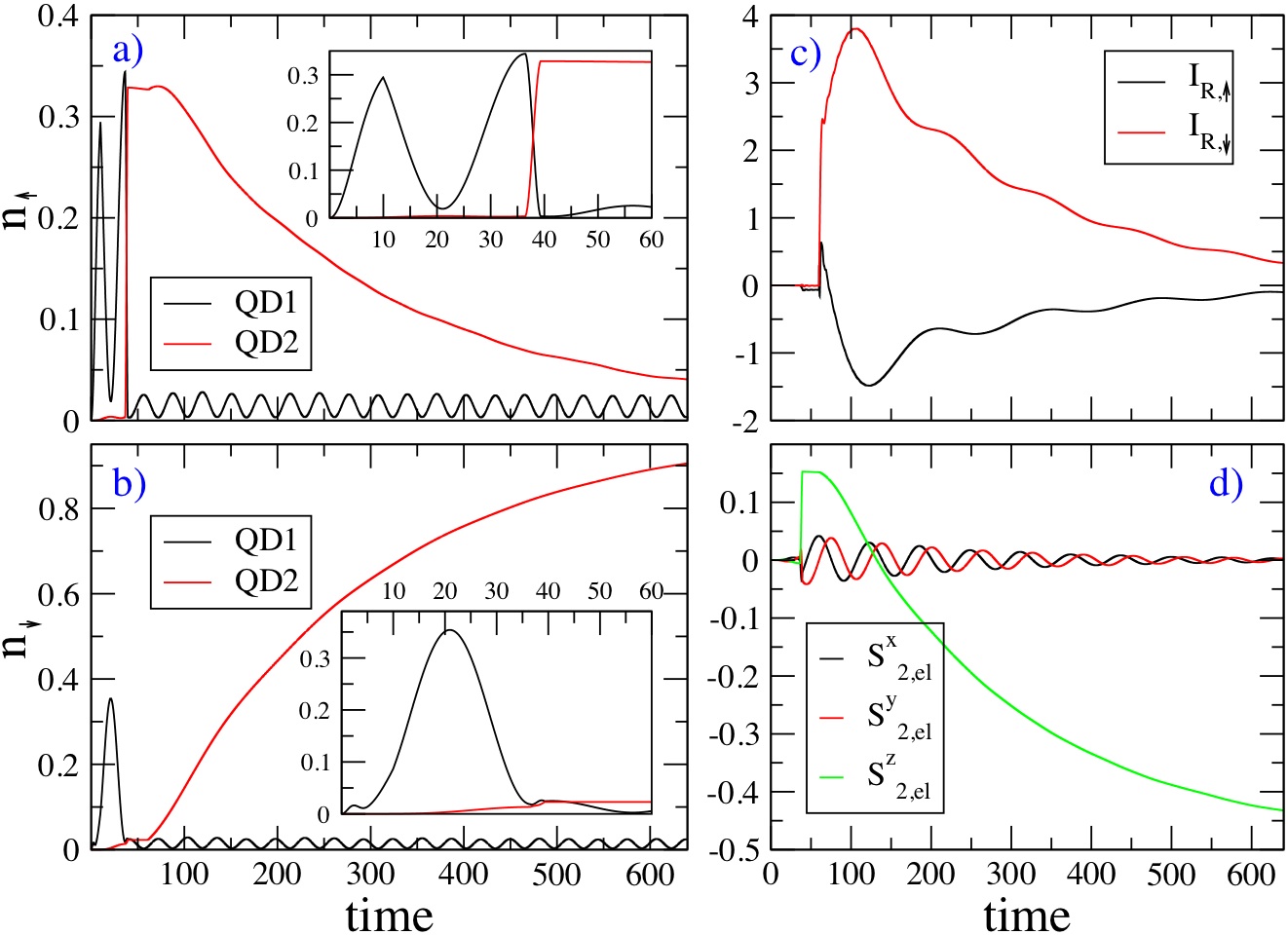}
\caption{Spin up density [panel a)] and spin down density [panel b)]
in QD1 and QD2 (the inset shows a magnification of the
time window $0<t<60$). The spin polarized current at the right
interface, $I_{R,\s}(t)$, is displayed in panel c) in units of $10^{-3}$, while panel d) shows
$\vec{S}_{2,\rm el}(t)$. For $t<T_{3}=39.32$ the
system parameters and the perturbations are the same as in Fig.
\ref{vqdsz} with $V_{\rm QD}^{(1)}=0.5$. For $t\geq T_{3}$ the
interdot hopping is lowered to the value $V_{\rm QD}^{(2)}=0.001$ and
QD2 becomes a well isolated system. At $t=60>T_{3}$ the hopping
$V_{R}(t)$ between QD2 and lead $R$ is raised up to $V^{(3)}=0.05$
and simultaneously a bias $U_{R,\ua}=U_{R,\da}=0.96$ is switched on
in the right lead. At this time the electrochemical potential in lead $R$ is
$\m_{R}=\ve_{\rm F}+0.96=0$ and lies in between the two spin levels of QD2.
}
\label{nupdown}
\end{figure}

In Fig. \ref{nupdown} we consider the same system parameters and
perturbations of Fig. \ref{vqdsz} (with $V_{\rm QD}^{(1)}=0.5$)
and fix the time $T_{3}=39.32$ when $S^{z}_{2,\rm el}$ has a
maximum. At $t= T_{3}$ the interdot hopping is lowered to $V_{\rm
QD}^{(2)}=0.001$ and QD2 becomes an almost isolated system. At
this stage the density of spin up and down electrons in QD2 is
practically constant as one can see from the insets of Fig.
\ref{nupdown} in panels a) and b). Shortly after $T_{3}$ the read
out phase starts. At $t=60$ we lower the barrier between QD2 and
lead $R$ and simultaneously switch on a bias
$U_{R,\ua}=U_{R,\da}=U_{R}=0.96$ in the right lead. The
electrochemical potential in lead $R$ becomes $\m_{R}=\ve_{\rm
F}+U_{R}=0$ and lies in between the two energy levels
$\ve_{2,\pm}=\pm J_{2}=\pm 0.05$ of the isolated QD2, with the
highest level $\ve_{2,+}$ for spin up electrons and the lowest
level $\ve_{2,-}$ for spin down electrons.\cite{note} Spin up
electrons in QD2 have, therefore, energy larger than $\m_{R}$ and
tunnel to the lead $R$. As a consequence the spin up density
decreases, as one can see in Fig. \ref{nupdown} panel a). On the
contrary, the lowest level $\ve_{2,-}$ has energy below $\m_{R}$
and a vanishingly small occupation. Spin down electrons tunnel
from lead $R$ to QD2 and the density of spin down electrons
increases, see Fig. \ref{nupdown} panel b). This charge transfer
generates a right-going spin-up current $I_{R,\ua}$ and a
left-going spin-down current $I_{R,\da}$, see Fig. \ref{nupdown}
panel c), which results in a large spin-current. The spin dynamics
in the $xy$ plane is displayed in Fig. \ref{nupdown} panel d)
where, besides the monotonically decreasing $z$ component, we plot
the $x$ and $y$ components of $\vec{S}_{2,\rm el}$. Due to the
symmetry of the problem $S^{x}_{2,\rm el}$ and $S^{y}_{2,\rm el}$
oscillate around zero with an exponentially decreasing amplitude.

\begin{figure}[htbp]
    \vspace{0.2cm}
\includegraphics*[width=.47\textwidth,height=5.5cm]{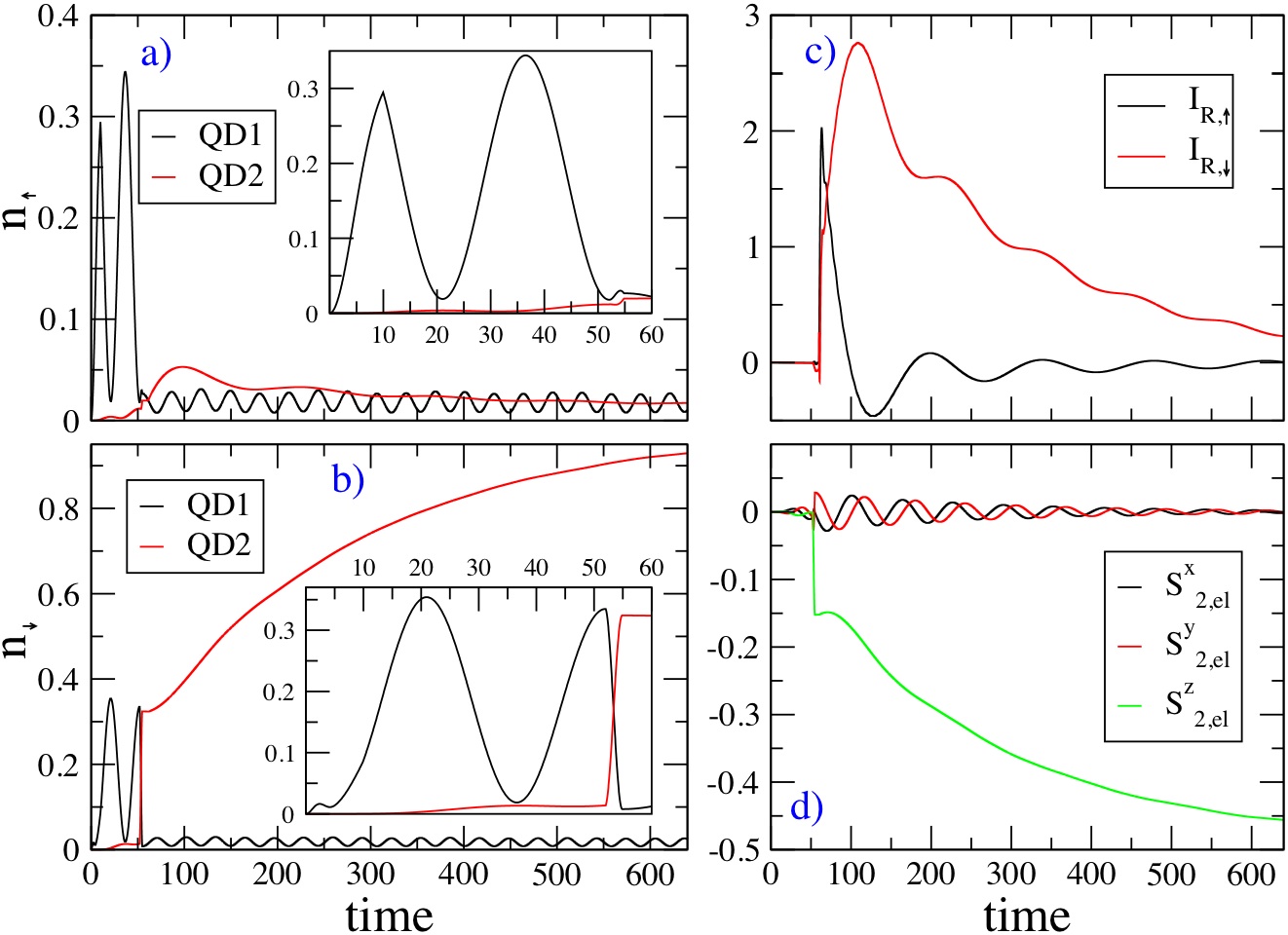}
\caption{Spin up density [panel a)] spin down density [panel b)] on
QD1 and QD2, spin polarized current at the right
interface, $I_{R,\s}$, in units of $10^{-3}$ [panel c)], and
$\vec{S}_{2,\rm el}$ [panel d)]. The insets of panels a)-b)
show a magnification of the density in the
time window $0<t<60$. Same system parameters and perturbations
of Fig. \ref{nupdown} but $T_{2}=52.05$ and $T_{3}=54.84$.}
\label{spindown}
\end{figure}

The situation corresponding to the antiparallel configuration in
QD2 is analyzed in Fig. \ref{spindown}. The difference with the
previous case is that we let the electron spin in QD1 rotate till is
polarized along the negative $z$ axis. The first time
$S^{z}_{1,\rm el}$ is minimum occurs at $T_{2}=52.05$, see insets
in panels a) and b). The spin transfer phase ends
at $T_{3}=54.84$ with an efficiency of about $90\%$. This can be seen
 in the inset of panel b)  where the spin down
density of QD2 swaps with that of QD1 in the time window $(T_{2},T_{3})$.
At $t=T_{3}$ the system undergoes the same perturbations considered
in Fig. \ref{nupdown}. Being the spin up level of QD2 scarcely
populated the change in the spin up density [panel a)]
and spin up current at the right interface [panel c)]
is very small as compared to the parallel configuration. A small change is
observed for the spin down quantities as well due to a population of
about 0.3 in the spin down level of QD2. Contrary to the parallel configuration set
up, the $z$ component $S^{z}_{2,\rm el}$ is negative when the read
out phase starts and does not change sign, see
panel d).

\begin{figure}[htbp]
    \vspace{0.2cm}
\includegraphics*[width=.47\textwidth,height=4.5cm]{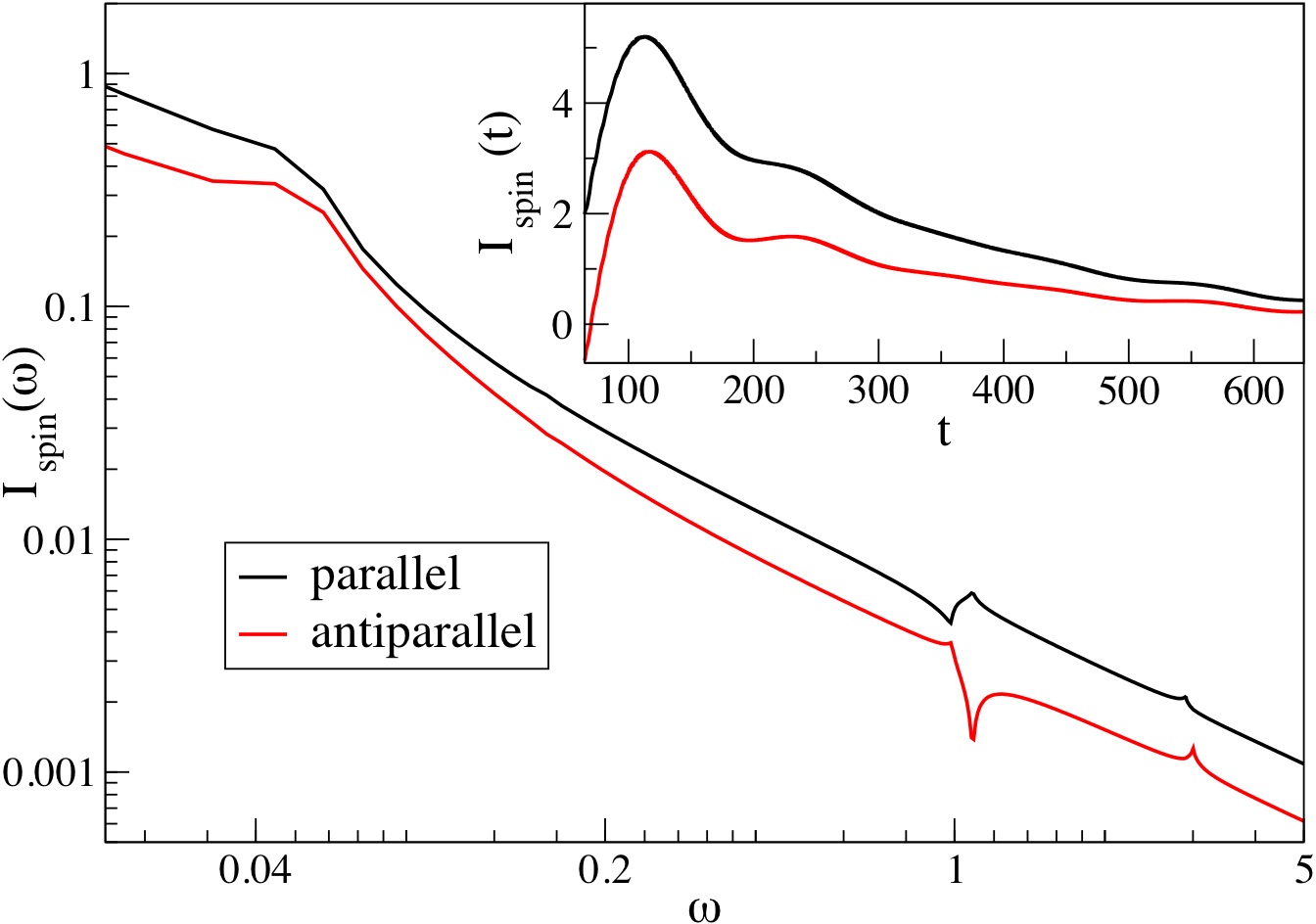}
\caption{Discrete Fourier transform of the
spin current $I_{\rm spin}=I_{R,\da}-I_{R,\ua}$
at the right interface for the parallel
and antiparallel configurations of Fig. \ref{nupdown} and
Fig. \ref{spindown} respectively. The inset shows  $I_{\rm spin}$ in
time domain, with $t$ in the range (65, 640), which has been
used to perform the discrete Fourier transform. Both
$I_{\rm spin}(\w)$ and $I_{\rm spin}(t)$ are in units of $10^{-3}$.}
\label{fourier}
\end{figure}

The spin current $I_{\rm spin}(t)\equiv I_{R,\da}(t)-I_{R,\ua}(t)$ during the
read-out phase ($t>60$) is displayed  in the inset of Fig.
\ref{fourier} for the parallel and antiparallel configurations
analyzed in Figs. \ref{nupdown}-\ref{spindown}. One
observes an exponential decay with superimposed oscillations of
frequency $|\ve_{\rm F}+U_{R}\pm J_{2}|=0.05$, as expected. However,
a closer inspection reveals a richer structure.
In Fig. \ref{fourier} we show the discrete Fourier transform of
$I_{\rm spin}(t)$ with $t$ in the range $(65,640)$. Besides the peak
at $\w=0.05$ there exist an extra peak at frequency $\w=|2-\ve_{\rm
F}|\sim 2.96$ and an asymmetric peak structure at frequency $\w=|\ve_{\rm
F}+2|\sim 1.04$. {\em The extra transient frequencies are due to the finite
bandwidth of the leads} since the energies +2 and -2 (in units of
$V$) correspond the top and the bottom of the right band.

\begin{figure}[htbp]
    \vspace{0.2cm}
\includegraphics*[width=.47\textwidth,height=5.cm]{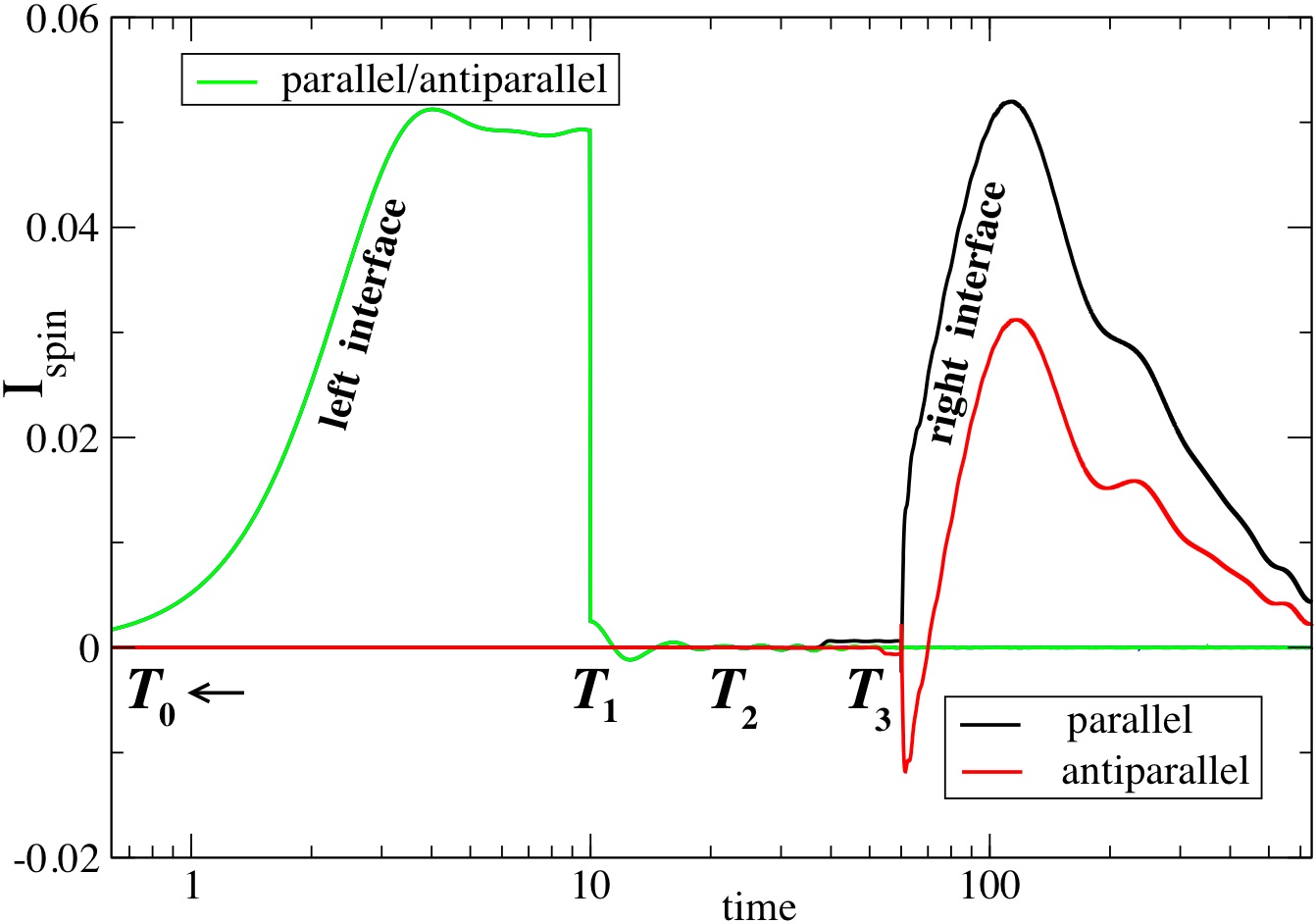}
\caption{Spin current at the left and right interfaces for the
parallel and antiparallel configurations of Fig. \ref{nupdown} and
Fig. \ref{spindown} respectively. For the parallel configuration
$T_{0}=0$, $T_{1}=10.0$, $T_{2}=36.5$, and $T_{3}=39.32$, while
for the antiparallel configuration $T_{0}=0$, $T_{1}=10.0$,
$T_{2}=52.05$, and $T_{3}=54.84$.} \label{spincurrbis}
\end{figure}

In conclusion, we have shown how to propagate in time a spinful
open quantum system subject to arbitrary time- and spin-dependent
perturbations. The semi-infinite nature of the leads has been
exactly accounted for. Full simulations of the microscopic charge
and spin transient dynamics of a double quantum dot in its
operating regime have been presented. Figure \ref{spincurrbis}
summarizes how the device works by displaying the spin currents at
the left and right interfaces during the entire sequence of
voltage pulses. Different processing of the injected spin up
current results in different spin currents at the right interface.

\section{Summary and Outlook}
\label{conc}

In the last few years we have witnessed an increasing interest
on transient responses in quantum transport mainly due to
their potential relevancy in molecular electronics, a field
where molecular devices will possibly operate under
non-steady-state conditions. The main difficulty in the study of
the short-time response of open quantum systems stems from the
macroscopic size of the leads. Several approaches have been proposed
to tackle this problem. Treating the leads in
the WBL approximation allows for obtaining a simple integral
equation for currents, densities,
etc.,\cite{jmw.1994,sl.1997.1,ylz.2000,zwymc.2007} but lacks
retardation effects. One-dimensional leads have been approximately
treated within a Wigner-function approach\cite{f.1990,dn.2005} or by including only a
finite number of lead unit-cells.\cite{ds.2006,as.2007,bsdv.2005}
Only recently it became possible to deal with the
semi-infinite nature of the leads using  a scheme based on
wave-functions propagation\cite{ksarg.2005} or, alternatively,
other algorithms based on solving the Dyson-Keldysh equations in
the time-domain.\cite{zmjg.2005,hhlkch.2006,mwg.2006,mgm.2007,hhlhc.2007}
Few attempts to include electron-correlation\cite{sl.1997.2,ktt.2001} as well as
electron-nuclear interactions\cite{vsa.2006,sssbht.2006} in the
transient regime have also been made.

In this work we have used a modified version of the propagation
algorithm of Ref. \onlinecite{ksarg.2005}, see Appendix \ref{alprop},
and generalized it to include the spin degrees of freedom. We have
proposed a double quantum dot system to manipulate the charge and the spin
of the electrons. Numerical
simulations of the entire operating regime have been provided. These
include some of the crucial steps in the theory of quantum
computation, like, e.g., the injection of spins and their read out.

The transient electron dynamics when a device is
perturbed by ultrafast voltage pulses is not only relevant to our
microscopic understanding but an exploitable feature to improve the
device performance. This has been explicitly shown in Section
\ref{numres}: the efficiency of the spin injection can be much higher
during the transient than at the steady state.
We also have found that for a given height of the barriers between
lead $L$ and QD1, QD1 and QD2, and QD2 and lead $R$, the damping of
the spin magnitude during the rotation phase is much smaller for
different exchange couplings, i.e., $J_{1}\neq J_{2}$, than for
$J_{1}=J_{2}$. This means
that the spin relaxation can be substantially
delayed using different quantum dots.

Using the non-equilibrium Green's function formalism in the WBL
approximation we have obtained a
rate equation for both the spin-injection and spin
read-out processes. For short times the rate equation becomes remarkably
transparent and permits us to identify the mechanisms leading
to a relaxation of the spin magnitude and to a deterioration of the
spin polarization. Going beyond the WBL approximation
results in a richer structure of the transient responses, as
transitions between the Fermi energy and the bottom/top of the band
occur as well.

As shown in Section \ref{numres}, the possibility of simulating
operational sequences like, e.g., that of Fig. \ref{schematic},
allows for a real-time study of fundamental processes not
accessible otherwise. Much more work is, however, needed before a
systematic comparison with experimental data can be made.
Accounting for intradot and possibly interdot electron-electron
interactions is of crucial importance for describing, e.g., the
Coulomb blockade or the Kondo regimes. The complications here stem
from the necessity of including electron correlations in a
time-dependent conserving manner, a progress which can be made
either within the framework of many body theory\cite{b.1962}  or
within one-particle frameworks like, e.g., time-dependent density
functional theory.\cite{rg.1984,vbdvls.2005} Developments in this
direction have been made in steady-state situations by treating
the correlation at the GW level.\cite{tr.2007,dfmo.2007,wshm.2008,tr.2008}

Another fundamental issue to be pursued is the extension to
three-dimensional leads. This would allow for a proper treatment
of the long-range Coulomb potential as well as for a realistic
description of the atomistic structure of the tunneling barriers.

Finally, the recent experimental advances in attaching quantum
dots to superconducting leads\cite{bosht.2007} prompt for a
generalization of the propagation algorithm to leads described by,
e.g., BCS-like models.
Such development
will give us access to a completely new phenomenology due to the
competition between the pairing interaction and the spin-flip interactions,
a topic not yet explored in the transient regime.

\acknowledgments

We would like to acknowledge useful discussions with S. Kurth. 
This work was supported in part by the EU Network of Excellence NANOQUANTA
(NMP4-CT-2004-500198).
E.P. is also financially supported by Fondazione Cariplo n. Prot. 0018524.

\appendix

\section{Proof of Eq. (\ref{ir})}
\label{proof}

The result in Eq. (\ref{ir}) is a consequence of the relative
orientation of the spin impurity $\vec{S}_{1}$ with respect to $\vec{S}_{2}$.
By definition the quantity $[\bgS_{2}^{z}(t)]_{1,1}$ is the (1,1) matrix element of
the product of three matrices
\begin{widetext}
\be
[\bgS_{2}^{z}(t)]_{1,1}=
\left[
e^{-it\left(\begin{array}{cc}
J_{1}\s_{x} & V_{\rm QD}\bfone_{2} \\
V_{\rm QD}\bfone_{2} & J_{2}\s_{z}
\end{array}
\right)}
\left(
\begin{array}{cc}
    \bfzero_{2} & \bfzero_{2} \\
    \bfzero_{2} & \s_{z}
\end{array}
\right)
e^{it\left(\begin{array}{cc}
J_{1}\s_{x} & V_{\rm QD}\bfone_{2} \\
V_{\rm QD}\bfone_{2} & J_{2}\s_{z}
\end{array}
\right)}\right]_{1,1} .
\label{ir2}
\ee
Consider the unitary operator $\bU=\bU_{g}\bU_{z}\bU_{x}$ which
consists of a rotation of both spin impurities around the $x$ axis by
an angle $\p$,
$
\bU_{x}=\left(
\begin{array}{cc}
    \exp[-i\p\s_{x}/2] & \bfzero_{2} \\
    \bfzero_{2} & \exp[-i\p\s_{x}/2]
\end{array}
\right),
$
followed by a rotation around the $z$ axis by an angle $\p$,
$
\bU_{z}=\left(
\begin{array}{cc}
    \exp[-i\p\s_{z}/2] & \bfzero_{2} \\
    \bfzero_{2} & \exp[-i\p\s_{z}/2]
\end{array}
\right),
$
followed by a gauge transformation which changes the sign of the
fermion operators on QD2,
$
\bU_{g}=\left(
\begin{array}{cc}
    \bfone_{2} & \bfzero_{2} \\
    \bfzero_{2} & -\bfone_{2}
\end{array}
\right).
$
Insertions of the identity matrix $\bU^{\dag}\bU$ in Eq. (\ref{ir2})
gives
\beq
[\bgS_{2}^{z}(t)]_{1,1}&=&
\left[\bU^{\dag}\bU
e^{-it\left(\begin{array}{cc}
J_{1}\s_{x} & V_{\rm QD}\bfone_{2} \\
V_{\rm QD}\bfone_{2} & J_{2}\s_{z}
\end{array}
\right)}\bU^{\dag}\bU
\left(
\begin{array}{cc}
    \bfzero_{2} & \bfzero_{2} \\
    \bfzero_{2} & \s_{z}
\end{array}
\right)\bU^{\dag}\bU
e^{it\left(\begin{array}{cc}
J_{1}\s_{x} & V_{\rm QD}\bfone_{2} \\
V_{\rm QD}\bfone_{2} & J_{2}\s_{z}
\end{array}
\right)}\bU^{\dag}\bU\right]_{1,1}
\nonumber \\
&=&-
\left[
e^{it\left(\begin{array}{cc}
J_{1}\s_{x} & V_{\rm QD}\bfone_{2} \\
V_{\rm QD}\bfone_{2} & J_{2}\s_{z}
\end{array}
\right)}
\left(
\begin{array}{cc}
    \bfzero_{2} & \bfzero_{2} \\
    \bfzero_{2} & \s_{z}
\end{array}
\right)
e^{-it\left(\begin{array}{cc}
J_{1}\s_{x} & V_{\rm QD}\bfone_{2} \\
V_{\rm QD}\bfone_{2} & J_{2}\s_{z}
\end{array}
\right)}\right]_{2,2}
=-[\bgS_{2}^{z}(-t)]_{2,2}.
\eeq
Taking into account that $[\bgS_{2}^{z}(t)]_{1,1}$,
$[\bgS_{2}^{z}(t)]_{2,2}$ are even functions of $t$, Eq. (\ref{ir})
follows.
\end{widetext}

\section{Propagation Algorithm}
\label{alprop}

Let $\bH(t)=\sum_{\a}\bH_{\a}(t)+\sum_{\a}(\bH^{0}_{\a
C}+\bH^{0}_{C\a})+\bH_{C}(t)$ be the one-particle Hamiltonian of a system which
consists of $\a=1,2,\ldots, N$ electrodes in contact with a central
region $C$. We assume that the time dependence of
\beq
\bH_{\a}(t)&=&
\left(\begin{array}{cc}
\bH_{\a,\ua }^{0} & \bfzero \\
\bfzero & \bH_{\a,\da}^{0}
\end{array}
\right)+
\left(
\begin{array}{cc}
    U_{\a,\ua}(t) &  \bfzero \\
    \bfzero & U_{\a,\da} (t)
\end{array}
\right)
\nonumber \\
&=&\bH_{\a}^{0}+\bU_{\a}(t)
\eeq
is a uniform spin-dependent
and time-dependent
shift while the time-dependence of $\bH_{C}$ has no restrictions.
We denote with $|\q_{\a}\ket$ the projection of a generic
wave-function $|\q\ket$ on electrode $\a$ and with $|\q_{C}\ket$
the projection of $|\q\ket$ onto region $C$. The time-dependent
Schr\"odinger equation reads
\beq
i\frac{\dr}{\dr t}|\q_{\a}(t)\ket=
\bH_{\a}(t)|\q_{\a}(t)\ket+
\bH^{0}_{\a C}|\q_{C}(t)\ket,
\label{se1}
 \\
i\frac{\dr}{\dr t}|\q_{C}(t)\ket=
\bH_{C}(t)|\q_{C}(t)\ket+\sum_{\a} \bH_{C\a}^{0}|\q_{\a}(t)\ket.
\label{se2} \eeq Performing the gauge transformation \be
|\q_{\a}(t)\ket=\exp[-i\int_{0}^{t}\dr\t \,
\bU_{\a}(\t)]\,|\f_{\a}(t)\ket, \ee and
$|\q_{C}(t)\ket=|\f_{C}(t)\ket$, Eqs. (\ref{se1}-\ref{se2}) become
\beq i\frac{\dr}{\dr t}|\f_{\a}(t)\ket=
\bH_{\a}^{0}|\q_{\a}(t)\ket+ \bH_{\a C}(t)|\f_{C}(t)\ket,
\label{2se1}
\\
i\frac{\dr}{\dr t}|\f_{C}(t)\ket=
\bH_{C}(t)|\f_{C}(t)\ket+\sum_{\a}
\bH_{C\a}(t)|\f_{\a}(t)\ket,
\label{2se2}
\eeq
with $\bH_{C\a}(t)=\bH_{C\a}^{0}\exp[-i\int_{0}^{t}\dr\t
\bU_{\a}(\t)]$
and $\bH_{\a C}(t)=[\bH_{C\a}(t)]^{\dag}$. The effect of the
gauge transformation is to transfer the time dependence from the
Hamiltonian describing the bulk electrodes to the Hamiltonian
describing the contacts between
the electrodes and region $C$. The gauge-transformed Schr\"odinger
equation is used to calculate the time evolved state $|\f(t_{m}=m\D
t)\ket\equiv |\f^{(m)}\ket$
by using the Cayley method
\be
(1+i\d \bH_{g}^{(m)})|\f^{(m+1)}\ket
=(1-i\d \bH_{g}^{(m)})|\f^{(m)}\ket,
\label{sec}
\ee
where $\d=\D t/2$,
$\bH_{g}^{(m)}=\frac{1}{2}[\bH_{g}(t_{m+1})+\bH_{g}(t_{m})]$,
and $\bH_{g}(t)=\sum_{\a}\bH_{\a}^{0}+\sum_{\a}(\bH_{\a
C}(t)+\bH_{C\a}(t))+\bH_{C}(t)$ is the gauge-transformed Hamiltonian.
The interface Hamiltonian $\bH_{\a C}$ is spin-diagonal provided
region $C$ includes the first few atomic layers of electrode $\a$. In
this case the projection of Eq. (\ref{sec}) onto different subregions
leads to a close recursive relation for the amplitudes $|\f_{C}^{(m)}\ket$
of the wave-function in region $C$ (the steps are similar to those of
Ref. \onlinecite{ksarg.2005})
\be
|\f^{(m+1)}_{C}\ket=
\frac{1-i\d \bH^{(m)}_{\rm eff}}{1+i\d \bH^{(m)}_{\rm eff}}|\f^{(m)}_{C}\ket
+|S^{(m)}\ket-|M^{(m)}\ket
\label{propf}
\ee
where the source term $|S^{(m)}\ket$ and the memory term
$|M^{(m)}\ket$ read
\be
|S^{(m)}\ket=\frac{-2i\d}{1+i\d \bH^{(m)}_{\rm eff}}
\sum_{\a}\bz_{\a}^{(m)}\bH_{C\a}^{0}
\frac{(1-i\d\bH^{0}_{\a})^{m}}{(1+i\d\bH^{0}_{\a})^{m+1}}|\f_{\a}^{(0)}\ket,
\ee
\beq
|M^{(m)}\ket&=&\frac{\d^{2}}{1+i\d \bH^{(m)}_{\rm eff}}\sum_{\a}
\sum_{j=0}^{m-1}\bz_{\a}^{(m)}
(\bQ_{\a}^{(j)}+\bQ_{\a}^{(j+1)})
\nonumber \\
&\times&
\bar{\bz}_{\a}^{(m-1-j)}(|\f_{C}^{(m-j)}\ket+|\f_{C}^{(m-1-j)}\ket).
\eeq
In the above equations we have used the following definitions
\be
\bz_{\a}^{(m)}=\frac{\exp[-i\int_{0}^{t_{m+1}}\dr\t \,
\bU_{\a}(\t)]+\exp[-i\int_{0}^{t_{m}}\dr\t \,
\bU_{\a}(\t)]}{2},
\ee
\be
\bQ_{\a}^{(m)}=\bH^{0}_{C\a}
\frac{(1-i\d\bH^{0}_{\a})^{m}}{(1+i\d\bH^{0}_{\a})^{m+1}}
\bH^{0}_{\a C},
\ee
\be
\bH^{(m)}_{\rm eff}=\bH_{C}^{(m)}-i\d\sum_{\a}\bz_{\a}^{(m)}
\bQ_{\a}^{(0)}\bar{\bz}_{\a}^{(m)}.
\ee
The recursive relation in Eq. (\ref{propf}) is written in terms of
matrices and vectors with the same dimension of the central region,
i.e., the infinitely large electrodes have been embedded in an
effective equation of finite dimension. We defer the reader to Ref.
\onlinecite{ksarg.2005} for the description of how to calculate the
matrices $\bQ_{\a}^{(m)}$ and the source term $|S^{(m)}\ket$.


\begin{thebibliography}{10}

\bibitem{nc.2000}
M. A. Nielsen and I. L. Chuang, {\em Quantum Computation and Quantum
Information}, Cambridge University Press, Cambridge, England, 2000.

\bibitem{asl.2002}
D. Awschalom, N. Samarth, and D. Loss, {\em Semiconductor
Spintronics and Quantum Computation}, Springer, Berlin,
2002.

\bibitem{ldv.1998}
D. Loss, and D. P. DiVincenzo,
Phys. Rev. A {\bf 57}, 120 (1998).

\bibitem{ms.2002}
{\em Spin-dependent transport in magnetic nanostructures},
edited by S. Maekawa and T. Shinjo,
Taylor and Francis, London and New York, 2002.

\bibitem{mba.2005}
S. Moskal, S. Bednarek, and J. Adamowski,
Phys. Rev. A {\bf 71}, 062327 (2005).


\bibitem{fhs.2006}
T. Fujisawa, T. Hayashi, and S. Sasaki,
Rep. Prog. Phys. {\bf 69}, 759 (2006).

\bibitem{slgj.2007}
F. M. Souza, S. A. Leao, R. M. Gester, and A. P. Jauho,
Phys. Rev. B {\bf 76}, 125318 (2007).

\bibitem{s.2007}
F. M. Souza,
Phys. Rev. B {\bf 76}, 205315 (2007).

\bibitem{hfcjh.2003}
T. Hayashi, T. Fujisawa, H. D. Cheong, Y. H. Jeong, and Y. Hirayama,
Phys. Rev. Lett. 91, 226804 (2003).

\bibitem{hwvwbek.2003}
R. Hanson, B. Witkamp, L. M. K. Vandersypen, L. H. Willems van Beveren, J.
M. Elzerman, and L. P. Kouwenhoven,
Phys. Rev. Lett. {\bf 91}, 196802 (2003).

\bibitem{ehvbwvk.2004}
J. M. Elzerman, R. Hanson, L. H. Willems van Beveren, B. Witkamp,
L. M. K. Vandersypen, and L. P. Kouwenhoven,
Nature (London) {\bf 430}, 431 (2004).

\bibitem{hvbvenkkv.2005}
R. Hanson, L. H. Willems van Beveren, I. T. Vink, J. M. Elzerman,
W. J. M. Naber, F. H. L. Koppens, L. P. Kouwenhoven, and
L. M. K. Vandersypen,
Phys. Rev. Lett. {\bf 94}, 196802 (2005).

\bibitem{pjtlylmhg.2005}
J. R. Petta, A. C. Johnson, J. M. Taylor, E. A. Laird, A. Yacoby,
M. D. Lukin, C. M. Marcus, M. P. Hanson, and A. C. Gossard
Science {\bf 309}, 2180 (2005).

\bibitem{kbtvnmkv.2006}
F. H. L. Koppens, C. Buizert, K. J. Tielrooij, I. T. Vink, K. C.
Nowack, T. Meunier, L. P. Kouwenhoven, and L. M. K. Vandersypen,
Nature (London) {\bf 442}, 766 (2006).

\bibitem{kstbfajfrp.2007}
M. Kataoka, R. J. Schneble, A. L. Thorn, C. H. W. Barnes, C. J. B. Ford,
D. Anderson, G. A. C. Jones, I. Farrer, D. A. Ritchie, and M. Pepper,
Phys. Rev. Lett. {\bf 98}, 046801 (2007).

\bibitem{cl.2007}
W. A. Coish and D. Loss,
Phys. Rev. B {\bf 75}, 161302(R) (2007).

\bibitem{kkr.2007}
A.V. Kimel, A. Kirilyuk, T. Rasing,
Laser Phot. Rev. {\bf 1}, 275 (2007)

\bibitem{zb.2002}
Jian-Xin Zhu and A.V. Balatsky,
Phys. Rev. Lett. {\bf 89}, 286802 (2002).

\bibitem{zc.2005}
H.-Q. Zhou, and S. Y. Cho,
J. Phys.: Condens. Matter {\bf 17}, 7433 (2005).

\bibitem{jsj.2001}
S. K. Joshi, D. Sahoo, and A. M. Jayannavar,
Phys. Rev. B {\bf 64}, 075320 (2001).

\bibitem{atz.2005}
A. Aldea, M. Tolea, and J. Zittartz,
Physica E {\bf 28}, 191 (2005).

\bibitem{cpz.2007}
F. Ciccarello, G. M. Palma, and M. Zarcone,
Phys. Rev. B {\bf 75}, 205415 (2007).

\bibitem{cpzov.2006}
F. Ciccarello, G. M. Palma, M. Zarcone, Y. Omar, and V. R. Vieira,
New J. Phys. {\bf 8}, 214 (2006).

\bibitem{cpzov.2007}
F. Ciccarello, G. M. Palma, M. Zarcone, Y. Omar, and V. R. Vieira,
J. Phys. A: Math. Theor. {\bf 40}, 7993 (2007).

\bibitem{ksarg.2005}
S. Kurth, G. Stefanucci, C.-O. Almbladh, A. Rubio and E. K. U. Gross,
Phys. Rev. B {\bf 72}, 035308 (2005).

\bibitem{c.1980}
M. Cini,
Phys. Rev. B {\bf 22}, 5887 (1980).

\bibitem{ls.2008}
H.-Z. Lu and S.-Q. Shen,
cond-mat/0804.1249v1.

\bibitem{sa.2004}
G. Stefanucci, and C.-O. Almbladh,
Phys. Rev. B {\bf 69}, 195318 (2004).

\bibitem{gksa.2001}
J. A. Gupta, R. Knobel, N. Samarth, D. D. Awschalom,
Science {\bf 292}, 2458 (2001).

\bibitem{mklsa.2007}
R. C. Myers, K. C. Ku, X. Li, N. Samarth, and D. D. Awschalom,
Phys. Rev. B {\bf 72},  R041302 (2005).

\bibitem{note}
The weak link $V^{(3)}=0.05$ between QD2 and lead $R$ yields two sharp
resonances at $\pm J_{2}$ in the local density of states of QD2.
Tuning the electrochemical potential
$\m_{R}=\ve_{\rm F}+U_{R}$ in between the resonances the
spin current is larger
as compared with the case $\m_{R}<-J_{2}$ or $\m_{R}>J_{2}$.

\bibitem{jmw.1994}
A.-P. Jauho, N. S. Wingreen, and Y. Meir,
Phys. Rev. B {\bf 50}, 5528 (1994).

\bibitem{sl.1997.1}
Q.-f. Sun and T.-h. Lin,
J. Phys.: Condens. Matter {\bf 9}, 3043 (1997).

\bibitem{ylz.2000}
J. Q. You, C.-H. Lam, H. Z. Zheng,
Phys. Rev. B {\bf 62}, 1978 (2000).

\bibitem{zwymc.2007}
X. Zheng, F. Wang, C. Y. Yam, Y. Mo, and G. H. Chen,
Phys. Rev. B {\bf 75}, 195127 (2007).

\bibitem{f.1990}
 W. R. Frensley,
 Rev. Mod. Phys. {\bf 62}, 745 (1990).

\bibitem{dn.2005}
Z. Dai, and J. Ni
Phys. Lett. A {\bf 342} 272 (2005).

\bibitem{bsdv.2005}
N. Bushong, N. Sai, and M. Di Ventra,
Nano Lett. {\bf 5}, 2569 (2005).

\bibitem{ds.2006}
A. Dhar, and D. Sen,
Phys. Rev. B {\bf 73}, 085119 (2006).

\bibitem{as.2007}
A. Agarwal, and D. Sen,
J. Phys.: Condens. Matter {\bf 19}, 046205 (2007).

\bibitem{zmjg.2005}
Y. Zhu, J. Maciejko, T. Ji, and H. Guo,
Phys. Rev. B {\bf 71}, 075317 (2005).

\bibitem{hhlkch.2006}
D. Hou, Y. He, X. Liu, J. Kang, J. Chen, and R. Han,
Physica E {\bf 31}, 191 (2006).

\bibitem{mwg.2006}
J. Maciejko, J. Wang, and H. Guo,
Phys. Rev. B {\bf 74}, 085324 (2006).

\bibitem{hhlhc.2007}
Y. He, D. Hou, X. Liu, R. Han, and J. Chen,
IEEE Trans. on Nanotech. {\bf 6}, 56 (2007).

\bibitem{mgm.2007}
V. Moldoveanu, V. Gudmundsson, and A. Manolescu,
Phys. Rev. B {\bf 76}, 085330 (2007).

\bibitem{sl.1997.2}
Q.-f. Sun, and T.-h. Lin,
J. Phys.: Condens. Matter {\bf 9}, 4875 (1997).

\bibitem{ktt.2001}
T. Kwapinski, R. Taranko, and E. Taranko,
Acta Physica Polonica A {\bf 99}, 293 (2001).

\bibitem{vsa.2006}
C. Verdozzi, G. Stefanucci, and C.-O. Almbladh,
Phys. Rev. Lett. {\bf 97}, 046603 (2006).

\bibitem{sssbht.2006}
C. G. Sanchez, M. Stamenova, S. Sanvito, D. R. Bowler, A. P. Horsfield, and T. N. Todorov,
J. Chem. Phys. {\bf 124}, 214708 (2006).

\bibitem{b.1962}
G. Baym,
Phys. Rev. {\bf 127}, 1391 (1962).

\bibitem{rg.1984}
E. Runge and E. K. U. Gross,
Phys. Rev. Lett. {\bf 52}, 997 (1984).

\bibitem{vbdvls.2005}
U. von Barth, N. E. Dahlen, R. van Leeuwen, and G. Stefanucci,
Phys. Rev. B {\bf 72}, 235109 (2005).

\bibitem{tr.2007}
K. S. Thygesen and A. Rubio,
J. Chem. Phys. {\bf 126}, 091101 (2007).

\bibitem{dfmo.2007}
P. Darancet, A. Ferretti, D. Mayou, and V. Olevano,
Phys. Rev. B {\bf 75}, 075102 (2007).

\bibitem{wshm.2008}
X. Wang, C. D. Spataru, M. S. Hybertsen, and A. J. Millis,
Phys. Rev. B {\bf 77}, 045119 (2008).

\bibitem{tr.2008}
K. S. Thygesen and A. Rubio,
Phys. Rev. B {\bf 77}, 115333 (2008).

\bibitem{bosht.2007}
C. Buizert, A. Oiwa, K. Shibata, K. Hirakawa, and S. Tarucha,
Phys. Rev. Lett. {\bf 99}, 136806 (2007).


\end{thebibliography}
\end{document}